# Dual-Guest Functionalised Zeolitic Imidazolate Framework-8 for 3D Printing White Light-Emitting Composites

*Abhijeet K. Chaudhari and Jin-Chong Tan\**

Multifunctional Materials & Composites (MMC) Laboratory, Department of Engineering Science, University of Oxford, Parks Road, Oxford, OX1 3PJ, United Kingdom.

*\*E-mail: jin-chong.tan@eng.ox.ac.uk*

## Abstract

Metal-organic frameworks (MOFs) stand as a promising chemically active host scaffold for the encapsulation of functional guests, because they could enhance luminescent properties by molecular separation of fluorophores in the nanoscale pores of the MOF crystals. Herein, we show simultaneous nanoconfinement of two fluorophores *viz*. (A) Fluorescein and (B) Rhodamine B in the sodalite cages of ZIF-8, constructed under ambient conditions through a simple one-pot reaction. We report a novel dual-guest@MOF system, termed: A+B@ZIF-8, which overcomes the intrinsic problem of aggregation caused quenching in the solid-state to gain bright yellow emission under UV irradiation. Subsequently, we combine this yellow emitter with a blue-emitting photopolymer resin, to yield a 3D printable luminescent composite material. We design a number of 3D printable composite objects for converting UV into warm white light emission, achieving a high quantum yield of ~44% in the solid-state 3D printed form. This research instigates the bespoke application of a vast range of 3D printable Guest@MOF designer composites targeting energy-saving lighting devices, smart sensors and future optoelectronics.

# 1. Introduction

World's 20% of energy is consumed by lighting on planet earth.[1-2] Consequently, substantial research in material science is focussed on the search for alternative solutions to yield more efficient solid-state lighting with enhanced quantum yield and emission lifetime.[2-3] In the pursuit of new materials, current research has concentrated on white light generation using luminescent organic molecules, polymers with red-green-blue (RGB) emitting side chains, three-layer RGB molecular system, hybrid perovskites, supramolecular polymers and lanthanide-based phosphor-converted white light-emitting diodes.[4-8] In the case of phosphor-converted white light-emitting diodes (LEDs), phosphors are composed of rare-earth elements that not only are of a limited supply, but also pose health hazards[9] and serious environmental impact.[10] In the light of this, there is an urgent need to develop and discover alternative materials to enable next-generation solid-state lighting applications.

The self-assembly of metal-organic frameworks (MOFs) or porous coordination polymers (PCPs) in an orderly fashion to form periodic framework structures offers several advantages.[11-13] More recently, researchers are exploring the use of periodic chemically-active molecular spaces found in the nanoscale pores of MOFs/PCPs (acting as a 'host' framework) to confine a range of functional 'guest' species or molecular complexes. The guest confinement approach yields a guest-host system termed 'Guest@MOF'[14] that can be useful as an electrical conductor,[15-16] light-emitting material,[17-19] selective gas/analyte sorbent, drug carrier, and chemical sensors.[20-21] For example, we have been working on a number of Guest@MOF systems to develop tuneable luminescent materials for photonics-based sensing applications.[22-24] In our studies, we found that the high concentration reaction (HCR) protocol is advantageous for: (i) rapid one-pot synthesis of nano-MOFs (e.g. nanosheets and nanostructures), (ii) *in situ* encapsulation of bulky guest molecules or nanoconfinement of metal complexes, and (iii) high yield of end products under ambient conditions.[22] These characteristics are the main requirements for constructing a functional material *via* facile guest confinement, with easy processability and future industrial-scale synthesis.

There is an increasing number of studies on Guest@MOF research reporting the encapsulation of a wide range of functional guests, such as rare-earth metals, organic dyes, carbon dots, quantum dots, and metal complexes to achieve tuneable luminescent properties and white light generation.[17-19, 25-32] Indeed, the majority of current efforts are much dependent on the use of non-economical and non-environmentally friendly rare-earth metals, see the statistics in Figure S1 of the Supporting Information (SI). Additionally, recent examples of synthetic method adopted for constructing tuneable luminescent systems require multistep core-shell processing to yield multivariate guest encapsulation,[33] which may be challenging to translate from lab to practical applications.

Herein we demonstrate the use of the HCR approach to accomplish the *in situ* multi-guest encapsulation of two fluorescent dye species (rare-earth free guests), being nanoconfined within the sodalite cage of the ZIF-8 framework *via* a one-pot reaction under ambient conditions. First, we present the facile synthesis and detailed characterisation of the previously unreported 'A+B@ZIF-8' compound (where, A: Fluorescein and B: Rhodamine B), which has a nanodisc morphology and emits yellow fluorescence (see Figure 1). Subsequently, we show how this new yellow-light emitter can be easily combined with a blue light-emitting photopolymer to form a composite that can be readily printed using a commercial 3D stereolithography machine. Finally, we demonstrate warm white light-emitting 3D printed objects and performed detailed characterisation of their photophysical properties central to the engineering of functional applications.

## 2. Results and Discussion

### 2.1 Rapid synthesis of dual-guest@MOF system

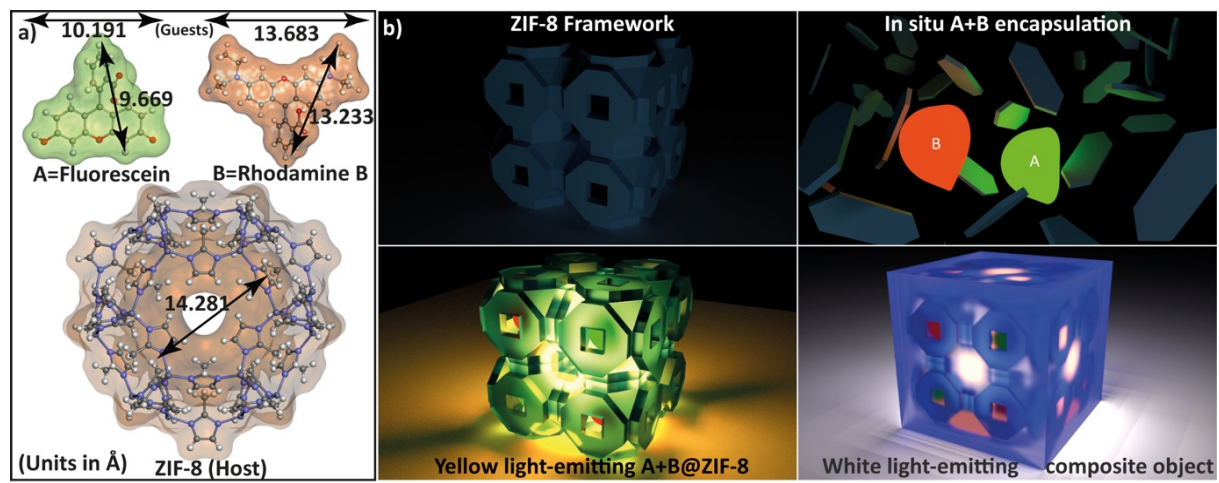

**Figure 1.** (a) Structural dimensions of the two fluorescent guest species A and B, and the host framework of ZIF-8 used in the current study. (b) Schematic representation of the concept adopted to fabricate A+B@ZIF-8 through *in situ* dual-guest encapsulation strategy, resulting in yellow-light emission in the solid state. Its subsequent use for fabricating a white-light emitting 3D object, which is a composite material of A+B@ZIF-8 combined with a blue light-emitting clear photopolymer resin.

On the one hand, we chose ZIF-8 as the host material because of its relatively large pore size within its sodalite cage (despite having small pore windows, see Figure 1a) together with its ease of synthesis. On the other hand, we chose Fluorescein (hereby denoted as 'A') and Rhodamine B (denoted as 'B') as the two fluorescent guest species for having strong green- and red-light emissions, respectively. Both the guests are ideal based on their molecular sizes and their ability to be constrained with the help of flexible/rotating groups like diethylamine (NEt$_2$) and phenyl-COOH in a nanoconfined environment (Figure 1a). Because of spatial constraints, it is possible for one dye species (A or B) to occupy a single pore cavity. The two different dye species, however, may be occupying adjacent pores of the assembled framework of the A+B@ZIF-8 compound. From the combination of both the dye molecules in methanol, we successfully achieved a deep-yellow colour emission by adjusting their volume ratios (Figure 2a inset). It was found that a combination of 25 mL 0.1 mM Fluorescein and 1.25 mL 0.1 mM Rhodamine B solutions exhibit a strong yellow light-emission when irradiated under the 365-nm UV source. This homogenous solution of A+B was used directly for constructing the 'Dual-Guest@MOF' system. Here, the combination of solutions was the important precursor for attaining solid-state yellow light-emitting A+B@ZIF-8 powder. The next crucial step was HCR synthesis of ZIF-8 by combining 3 mmol (0.8923g) of $Zn(NO_3)_2$ in 3 mL of methanol and 7.5 mmol (0.61575g) of deprotonated 2-methylimidazole (mIm) in 3 mL of methanol (mIm was deprotonated using 7.5 mmol in 1.04 mL of trimethylamine, $NEt_3$) in the presence of the A+B methanolic solution.

By employing the HCR method,[22] we obtained a readily forming product as soon as the linker and metal ion solutions were mixed together in the presence of the A+B guest solution. The product formed was thoroughly washed using methanol (where both the A and B guests, as well as host reactants are highly soluble) to remove non-confined guest molecules, and the excess mIm and $Zn^{2+}$ reactants. Further details of the synthesis are given in the SI (sections 2 and 3). Notably, the wet sample after washing and its dried powder have retained the yellow emission evidenced in the A+B solution combination (Figure 2a). Here, we note that the ZIF-8 host plays an important role by separating the single molecules from each other, thereby maintaining the solution-like optical properties that are generally quenched in the solid-state form of pure emitters (like A and B employed in the current study, see Figure S2). To confirm the successful framework formation and A+B guest confinement within it, material characterisation was performed. Powder X-ray diffraction (XRD) of the dry final product confirmed the formation of A+B@ZIF-8, without major changes to the original structure of ZIF-8, though some broadening of the Bragg peaks was noticed (Figure 2a). Morphological characterisation of the compound using atomic force microscopy (AFM) revealed the nanodisc-like morphology with an aspect ratio ranging from 5:1 to 10:1, where each nanodisc exhibits a thickness of under ~20 nm with a lateral dimension of over 100~200 nm (Figure 2b, and representative AFM images in Figures S5). The presence of nanostructures supports the broadening of XRD peaks observed in Figure 2a.

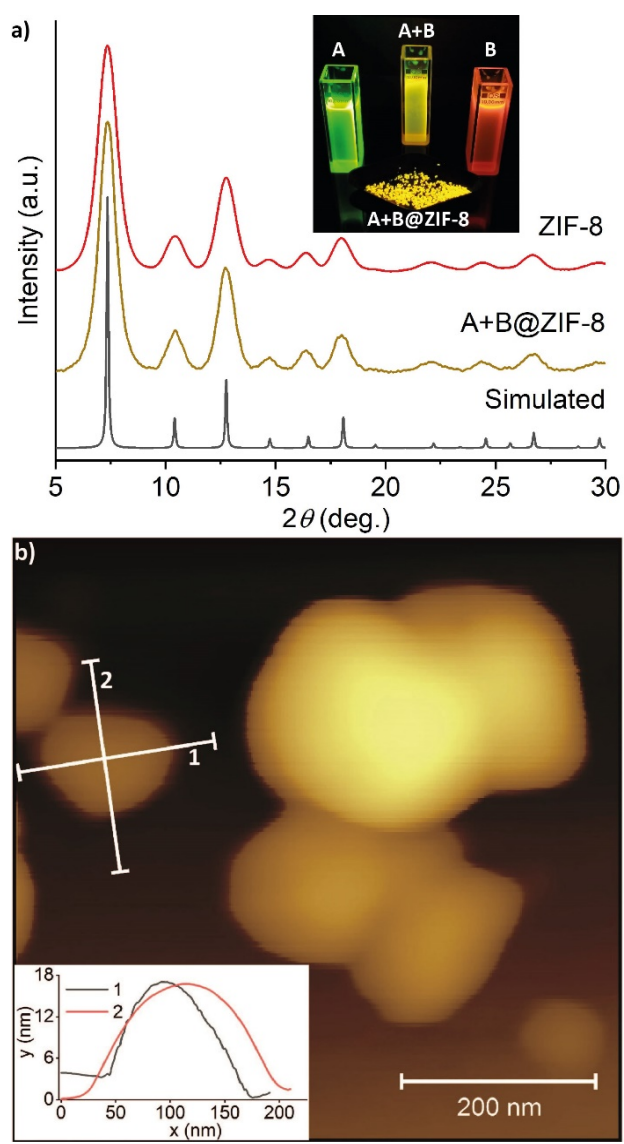

**Figure 2.** (a) X-ray diffraction of A+B@ZIF-8 powder confirming the crystal structure of ZIF-8 obtained by one-pot high-concentration reaction method. Inset shows the emissions observed under a 365 nm UV lamp: solid-state A+B@ZIF-8 powder and solution-state A+B combination, as compared with the pure solutions of (A) Fluorescein and (B) Rhodamine B in methanol. (b) Atomic force microscopy image of the A+B@ZIF-8 crystals exhibiting the nanodisc morphology, whose aspect ratio (length/height) is about 10:1 as shown in the inset.

## 2.2 3-D printing of light-emitting composite objects

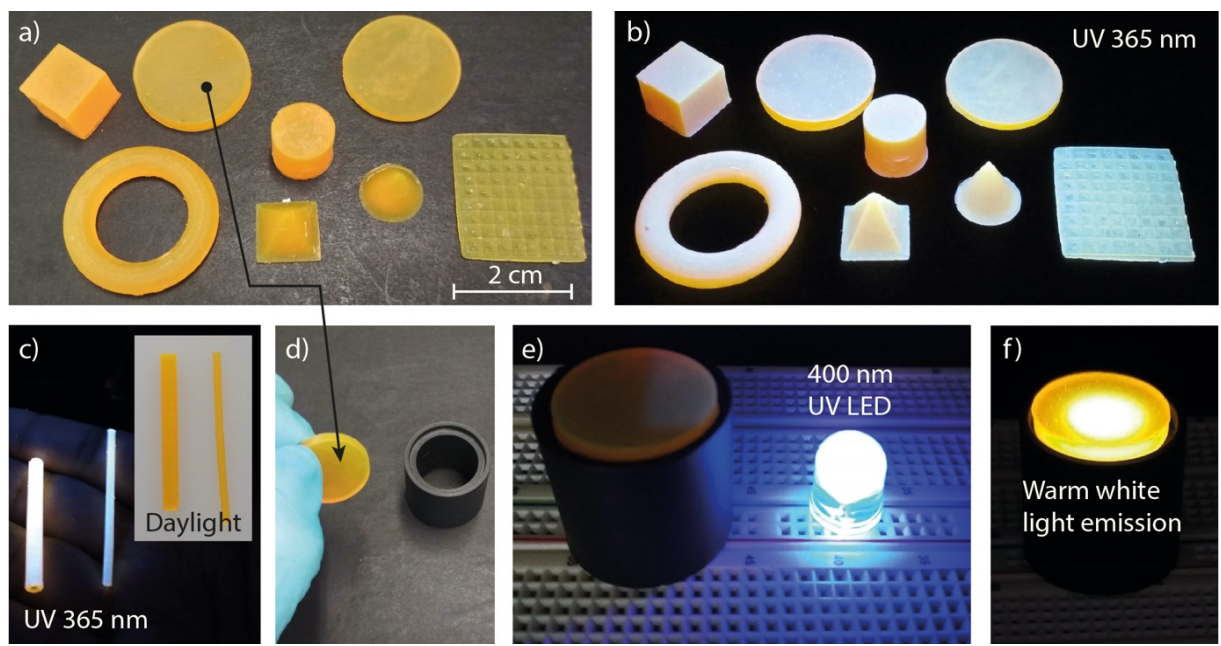

**Figure 3.** (a) Various shapes of 3D-printed objects constructed from the composite made of A+B@ZIF-8 powder dispersed in a clear photopolymer resin. (b-c) The 3D-printed composite produces white light emission under UV irradiation. (d) A 3-mm thick 3D printed disc-shaped pellet installed on a black cylindrical support (e) for converting 400 nm UV irradiation from an LED to generate (f) warm white-light emission.

Using the A+B@ZIF-8 powder, now we proceed with the creation of 3D-printed objects capable of generating a white light emission. A commercially available optically transparent photopolymer with blue emission under UV ('clear resin' by Formlabs, see SI section 4) was used for this purpose. We established that a combination of 10 g of yellow light-emitting A+B@ZIF-8 compound with 50 mL of blue light-emitting clear resin results in a stable white light-emitting composite after printing (Figure 3). The main strategy implemented here for obtaining a homogenous mixture of A+B@ZIF-8 and clear resin was the use of a wet sample of the former after thorough washing and centrifugation steps. The mixture was stirred together at room temperature for ~12 hours in the dark, to achieve homogeneity of the mixture while avoiding curing of resin by the surrounding light (see Figure S3). This composite mixture stayed consistently homogenous after overnight mixing without visible agglomeration of solid particles. The mixture was directly used for printing a wide range of 3D objects shown in Figure 3a-c. In Figures 3d-f, we demonstrated a 3-mm thick 3D printed composite pellet for generating warm white light when exposed to a 400 nm UV LED. Colour chromaticity diagram in Figure 4 shows that the 3D printed composite (A+B@ZIF-8 in clear resin) could convert UV light, where the CIE coordinates were determined to change from ($x$, $y$) values of (0.1888, 0.1114) to (0.4117, 0.4790) hence shifting to a warm white light emission (defined in the range of about 3100-4500 K). Furthermore, we demonstrate that it is straightforward to tune the emission chromaticity to cover a broad range of colour temperatures, from cool to warm white light, by systematically changing the thickness of the 3D printed pellets. Figure 4b shows the evolution of the emission spectra as a function of pellet thickness. The systematic shift of CIE coordinates when varying the pellet thickness between 0.5 mm and 5.2 mm is demonstrated in Figure 4(c).

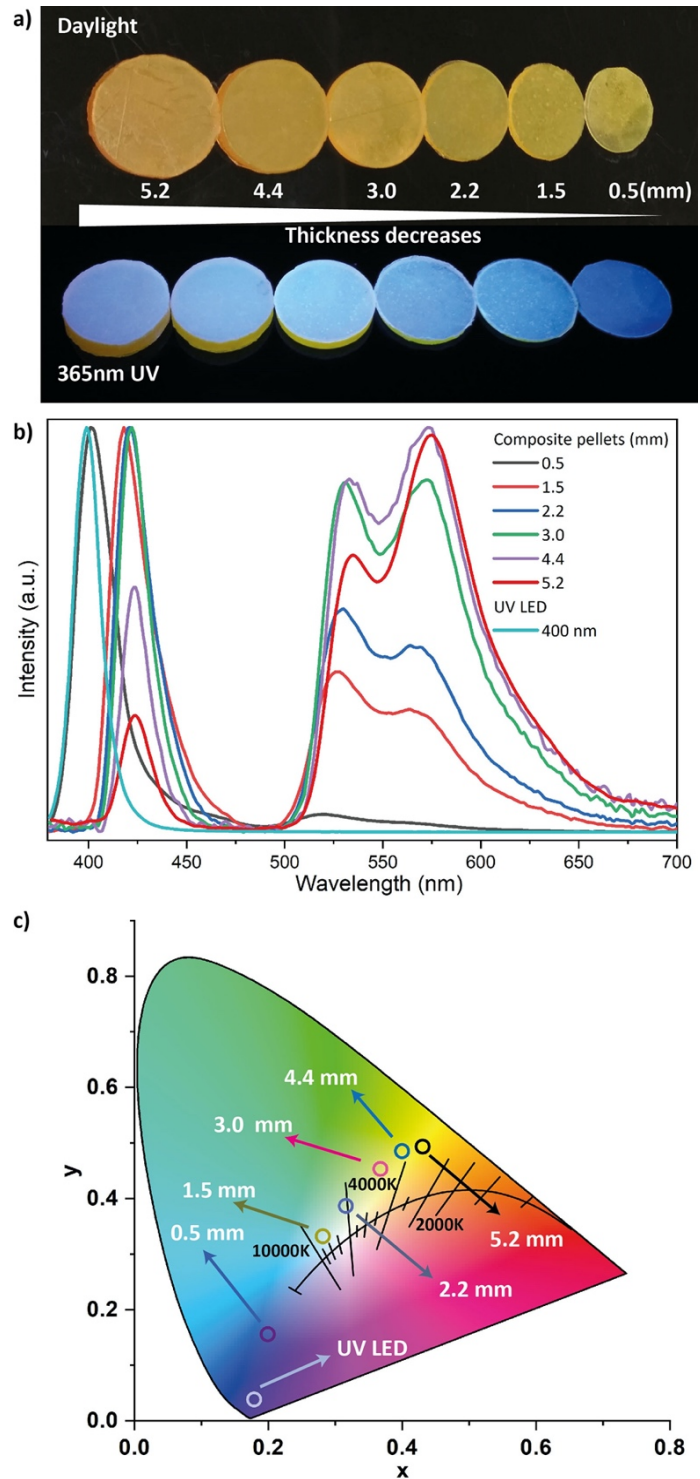

**Figure 4.** (a) Photograph taken under the 365-nm UV lamp, showing the emission of the 3D printed pellets of different thicknesses. (b) Tuneable pellet emission when being employed as the lid of a light conversion device (Figure 3f) and irradiated by a 400-nm UV LED. Note that the passage of the 400-nm UV source through the clear lid has been blocked by a thickness of between 0.5-1.5 mm. (c) CIE coordinates and colour

temperature showing the effects of pellet thickness, showing the ability for tuning the emission ranging from cool white light to warm white light.

**2.3 Photophysical properties**

We have investigated in detail the optical properties of the A+B@ZIF-8 compound and the resulting 3D-printed composites to understand the effects of confinement of fluorescent guests by the framework of ZIF-8, and when it is incorporated within the photopolymer resin. From the beginning we observed the bright emission arising from the solid-state powder of A+B@ZIF-8, which is a clear indication of single molecule separation of emissive guest species by the nanopores of ZIF-8. This observation was later confirmed by contrasting the solution-state and solid-state optical properties data shown in Figure 5.

Figure 5a shows the excitation spectra of the solutions and solids. Solution A (Fluorescein) has excitation maximum at 494 nm; at 496 nm in combination with solution B (Rhodamine B), and at 511 nm in A+B@ZIF-8 solid-state form as well as in the composite pellet form. Turning to solution B, it exhibits excitation maximum at 544 nm but retains this wavelength in combination with solution A; at 555 nm in A+B@ZIF-8 solid-state form, and at 554 nm in the composite pellet form.

Figure 5b compares the emission spectra of the solutions and solids. When excited, the emission maximum of solution A was observed at 515 nm; at 516 nm in combination with solution B, however it red shifted to 542 nm for A+B@ZIF-8 solid-state form and again blue shifted to 528 nm for the composite pellet form. In the case of solution B, emission maximum was observed at 570 nm, it maintained the same position in combination with solution A, but red shifted to 580 nm for A+B@ZIF-8 solid state form and again blue shifted to 572 nm for the composite pellet form. The emission spectra of the composite pellet are independent of the excitation wavelengths (Figure S5).

The modification to the photophysical properties of the fluorescent guest and the host framework discussed above reveal the effects of guest-host interaction. The major interacting groups at the ground state located around the environment of guest species are the -N-Zn-N- linkages and the H-C=C-H group of mIm, which can polarise A and B to alter their optical properties and resulting in the red shifted emission. For the 3D printed composite pellet, we reasoned that the observed blue shifted emission is due to the structural relaxation of A+B@ZIF-8 upon its dispersion in the resin photopolymer. Both guest species, A and B possess hydrogen bond-making and π-π stacking phenyl-COOH group as well as the 'O' and 'N$^+$' (for B) sites, which could easily interact with the surrounding framework environment. Moreover, the emission spectrum of the clear resin photopolymer exhibits emission maximum at 440 nm in the 3D-printed blank pellet versus 434 nm in the composite pellet. Interestingly, this means that the emission spectrum of resin has a strong overlap with the excitation spectra of A and B (Figure 5c), hence facilitating the excited-state energy transfer from the clear photopolymer resin to the confined guest emitters. For further comparison the diffuse reflectance spectra of the 3D printed pellet and its constituents in the solid state are shown in Figure S7, showing the overlapping absorption bands associated with host-guest interactions.

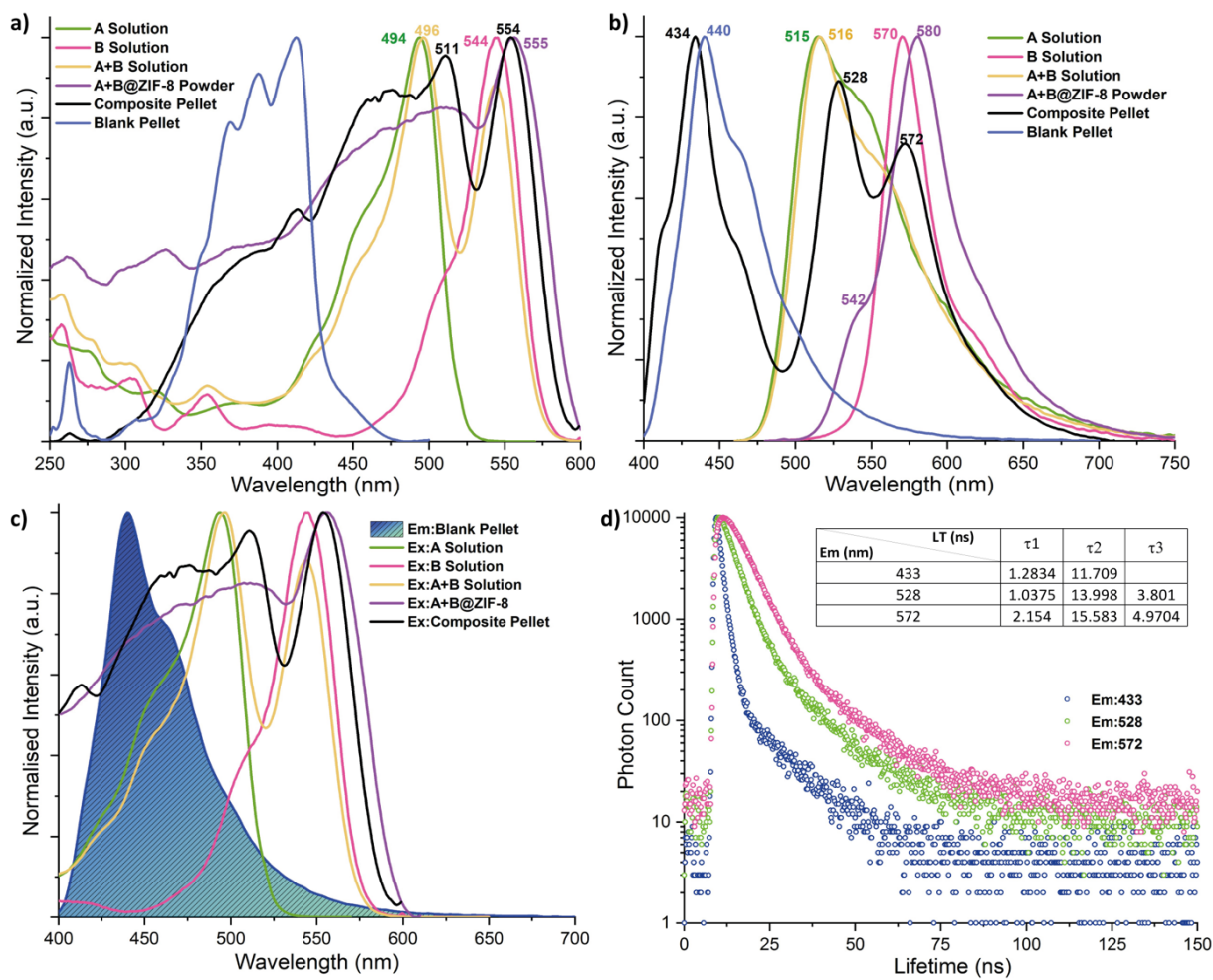

**Figure 5.** (a) Excitation and (b) emission spectra of the A and B guest species in solution form (isolated or combined), A+B@ZIF-8 in solid-state form (dual-guest@MOF powder), 3D printed blank pellet (photopolymer clear resin) and composite pellet (A+B@ZIF-8 combined with clear resin). (c) Emission spectrum of 3D printed blank pellet compared to the excitation spectra of other fluorescent species to identify overlap for energy transfer in excited state. (d) Fluorescence lifetime decay curves of the white light-emitting composite pellet measured at three different emission wavelengths, the table inset summarises the time constants ($\tau$) in nanoseconds.

The energy transfer process was studied by fluorescence lifetime (FLT) measurements employing the time-correlated single photon counting (TCSPC) technique. Decrease in FLT with energy transfer from donor to acceptor is indicative of a non-radiative process as evidenced from FLT of the clear resin (blank pellet), which fell from 15.40 ns to 13.99 ns (for emission at 528 nm), and to 11.70 ns (for emission at 433 nm). No major change in FLT of resin was observed for emission at 572 nm (see more details in Table S1). A short lifetime of ~2 ns corresponds to emission of the mIm linker of ZIF-8, the lifetime greater than 11 ns corresponds to the resin, while the lifetimes of 3.801 ns and 4.9704 ns can be ascribed to guest A and guest B, respectively. Detailed FLT data and curve fitting results are given in Figures S8-S17 in the SI. The quantum yield (QY) values of the A+B@ZIF-8 solid powder and the 3D-printed composite pellet are 47.3% and 43.6%, respectively (Table S2). In fact, their QY values are appreciably higher versus other solid-state Guest@MOF luminescent materials found in the literature, for example: QY = 30% for Zr-NDC MOF,[30] QY = 17.4% for ZJU-28,[28] and QY = 20.4% for iridium-complex@MOF.[17]

## 3. Conclusions

In summary, we show for the first time a Dual-Guest@MOF material which can be 3D printed to construct a warm white light-emitting device. In fact, the colour chromaticity of the device can be straightforwardly adjusted from a cool white light to a warm white light emission (CCT: 8300 K → 3700 K), simply by varying the thickness of the pellets. Our HCR synthetic protocol is versatile, as it enables simultaneous nanoscale confinement of multivariate fluorescent guest species within the MOF pores. Significantly, the separation of fluorescent dye molecules prevents the aggregation caused quenching phenomenon, therefore enabling bright light emission to be retained in the solid state. The facile one-pot synthetic strategy we demonstrated here is easy to implement and has a good yield in terms of materials production. It can be tailored for large-scale fabrication of fluorescent materials and adapted to confine multiple fluorescent dyes aimed at bespoke lighting applications. It is important to recognise that the direct addition of dye molecules into resin could not yield the same emission as obtained from the dispersion of A+B@ZIF-8 in resin, because the dilution effect of unencapsulated dyes could alter the resultant emission spectra. Finally, we show the efficacy for combining fluorescent Guest@MOF compounds with a photopolymer resin to enable 3D printing of a wide range of geometries and designs, as such opening the door for the engineering of photonic sensors, optoelectronics, and future metamaterials. Our study therefore demonstrates the advantages of employing the Dual-Guest@MOF approach: on the one hand, to permit chromaticity tuning, and, on the other hand, for enhancing structural and photostability of functional devices *via* the manufacturing of 3D printed composite objects.

**Acknowledgements**

We thank the ERC Consolidator Grant under the grant agreement 771575 (PROMOFS) for supporting the research. We are grateful to the Research Complex at Harwell (RCaH) for access to the materials characterization facilities.

**Conflicts of interest**

There are no conflicts to declare.

*Supporting Information*

*for*

# Dual-Guest Functionalised Zeolitic Imidazolate Framework-8 for 3D Printing White Light-Emitting Composites

*Abhijeet K. Chaudhari and Jin-Chong Tan\**

*Multifunctional Materials & Composites (MMC) Laboratory,*
*Department of Engineering Science, University of Oxford,*
*Oxford OX1 3PJ, United Kingdom*

*[jin-chong.tan@eng.ox.ac.uk](mailto:jin-chong.tan@eng.ox.ac.uk)

**Table of Contents**

# 1 Statistics on luminescent MOF research

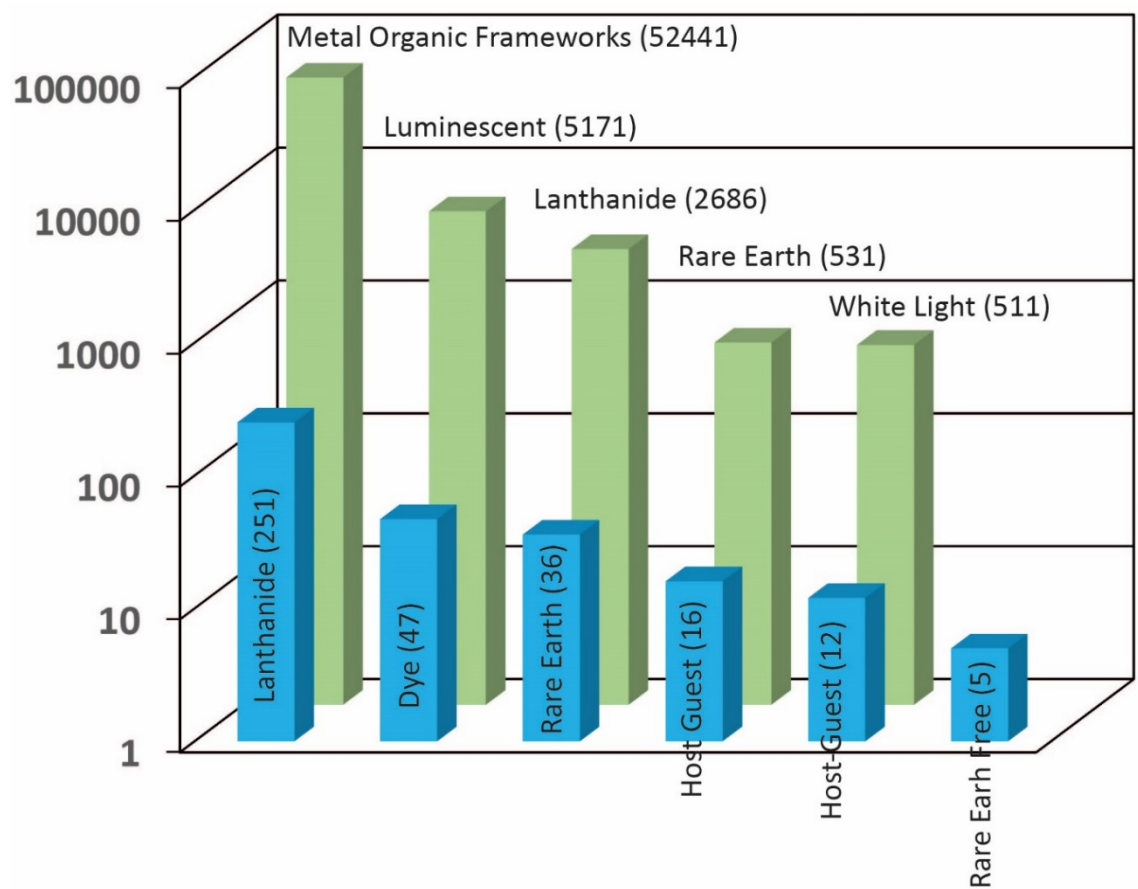

**Figure S1**. Statistical data indicating number of publications related to luminescent properties of MOF and MOF-derived materials under the main topic of metal organic frameworks. The statistics above are direct search results on the Web of Science (25/06/2019) using the keywords specified on the bars. Blue bars represent MOF-based white-light emitters.

## 2  Synthesis of yellow light-emitting solution

Yellow light-emitting solution was made by combining 0.1 mM solution of Rhodamine B and 0.1 mM solution of Fluorescein in methanol solvent. Initially 25 mL of Fluorescein was taken, into which Rhodamine B solution was slowly added to yield the yellow emission. The final combined solution has a volumetric ratio of 25:1.25 mL for Fluorescein to Rhodamine B, respectively. The ratio was determined by stepwise addition of 300 μL of solution B into a 25 mL of solution A. At each increment point, the change in emission was monitored under 365 nm UV lamp employing the UPRTek handheld photospectrometer. In this work, the final volume of B for addition into A to yield a warm white light emission was determined by examining the variation of the colour chromaticity using the CIE diagram in Figure S2.

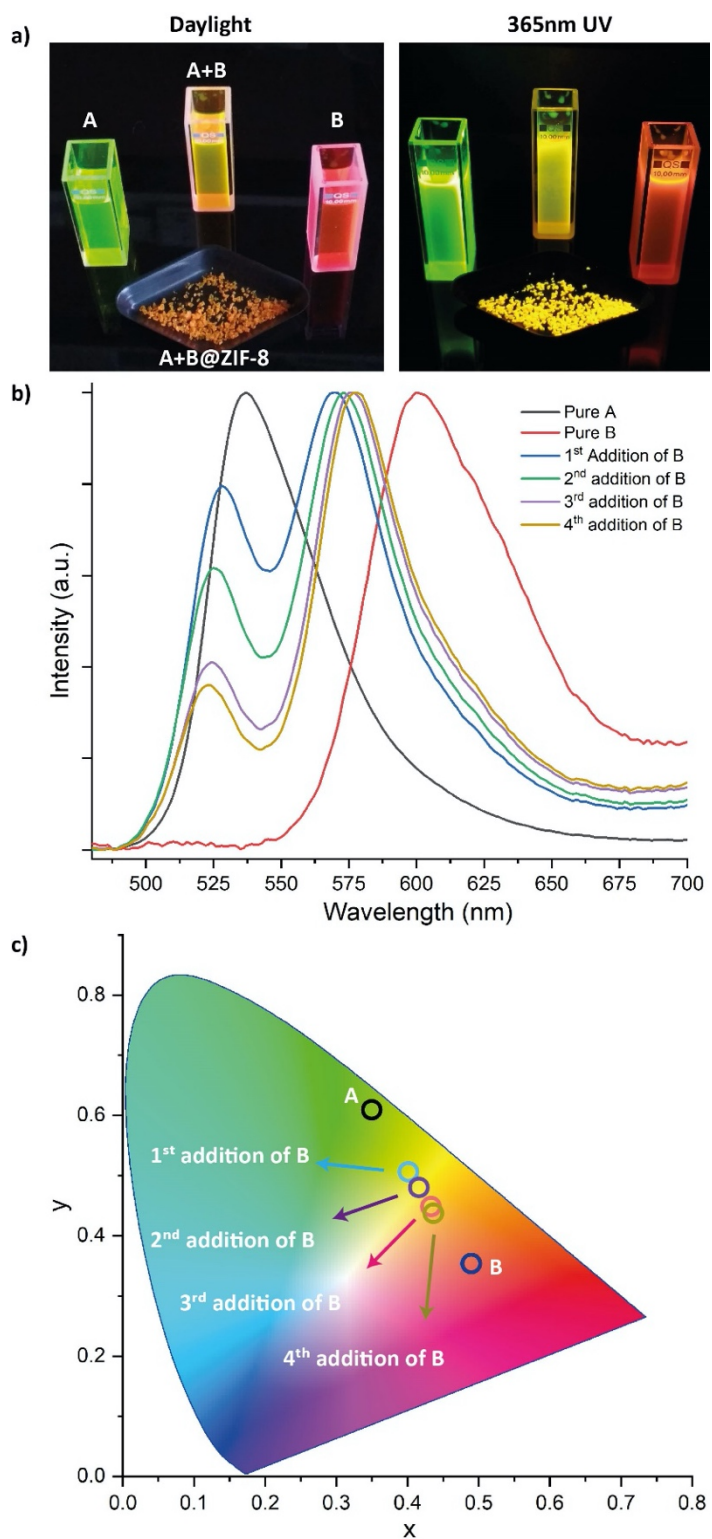

**Figure S2**. (a) Photograph taken under the 365 nm UV lamp, showing the emission of solutions A = Fluorescein, B = Rhodamine B, A+B mixture, and the solid-state emission of A+B@ZIF-8 powder (after drying). (b) Changes in the emission spectra of the combined solutions of A+B upon addition of 300 μL of B into A at each incremental point, (c) CIE 1931 diagram showing changes in colour chromaticity corresponding to the spectra in (b).

## 3 Synthesis of yellow light-emitting A+B@ZIF-8 powder

Yellow light-emitting solid powder was made by using the yellow solution of A+B described in §2 above, Zn(NO$_3$)$_2$ and 2-methylimidazole (mIm). Two separate solutions were prepared: (i) 3 mmol of Zn(NO$_3$)$_2$ in 3 mL methanol to yield a Zn$^{2+}$ solution, and (ii) 9 mmol of mIm (deprotonated using 9 mmol of trimethylamine) in 3 mL methanol.

In the encapsulation reaction, first, the A+B yellow solution was mixed with Zn$^{2+}$ solution (i). Then, a deprotonated solution of mIm (ii) was quickly added to (i), which immediately formed a thick precipitated product. The precipitate was then thoroughly washed 3 times with methanol, and the product isolated by centrifuge and dried. During each washing cycle, the material was subjected to 50 mL of methanol and 10 minutes of sonication to separate the aggregates. After sonication, the solid product of A+B@ZIF-8 was separated by centrifugation for 10 minutes at 8000 rpm, and then dried at 70 °C for 4 hr.

In contrast to the approach implemented above, we have also attempted to yield white light emission using the separate scheme of A@ZIF-8 and B@ZIF-8 *via* two separate synthesis and subsequent combination in resin to tune the emission. However, this approach failed due to inhomogeneity problem and non-uniformity in emission of the 3D printed (bulk) sample. Therefore the dual-guest approach has the advantage compared to the (conventional) individually encapsulated MOF approach.

## 4  3D printing of A+B@ZIF-8 dispersed in a clear photopolymer resin

3D printing was conducted using the Form 2 stereolithography printer made by Formlabs. For printing the 'blank' pellets, we used the clear resin commercially available from Formlabs, without further alteration. This clear resin is a photopolymer mixture of methacrylic acid esters and photoinitiator, designed for curing using the 405 nm laser of the Form 2 printer. However, the 'composite' pellets were 3D printed from the homogenous mixture of 10 g of A+B@ZIF-8 combined with 50 mL of clear resin. To obtain a homogenous composite mixture of resin and MOF powder, we used wet powder sample (i.e. after washing and centrifugation, but before drying) to mix with clear resin to minimise aggregation. To improve MOF dispersion in the clear resin, the mixture was stirred in the dark for ~12 hours using a mechanical stirrer, see Figure S3 (for protection against any visible light that might partially cure the resin). Subsequently, the 3D printing process was carried in automatic mode where the composite resin was dispensed from a 1-L cartridge into a standard resin tank (~130 mL), equipped with temperature control (31.5 °C) and automatic wiper.

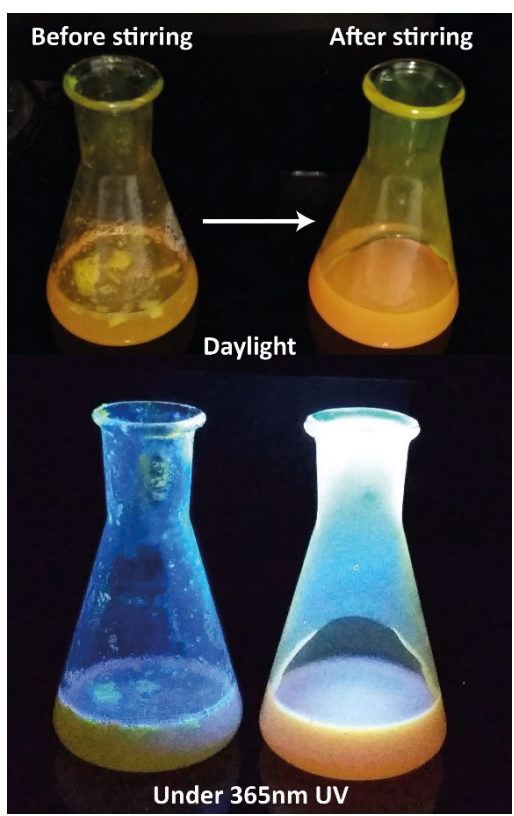

**Figure S3**. Photograph showing the effect of overnight stirring of (wet powder) A+B@ZIF-8 into the clear photopolymer resin, it was found that the non-homogenous mixture turned into a uniform homogenous suspension after overnight stirring.

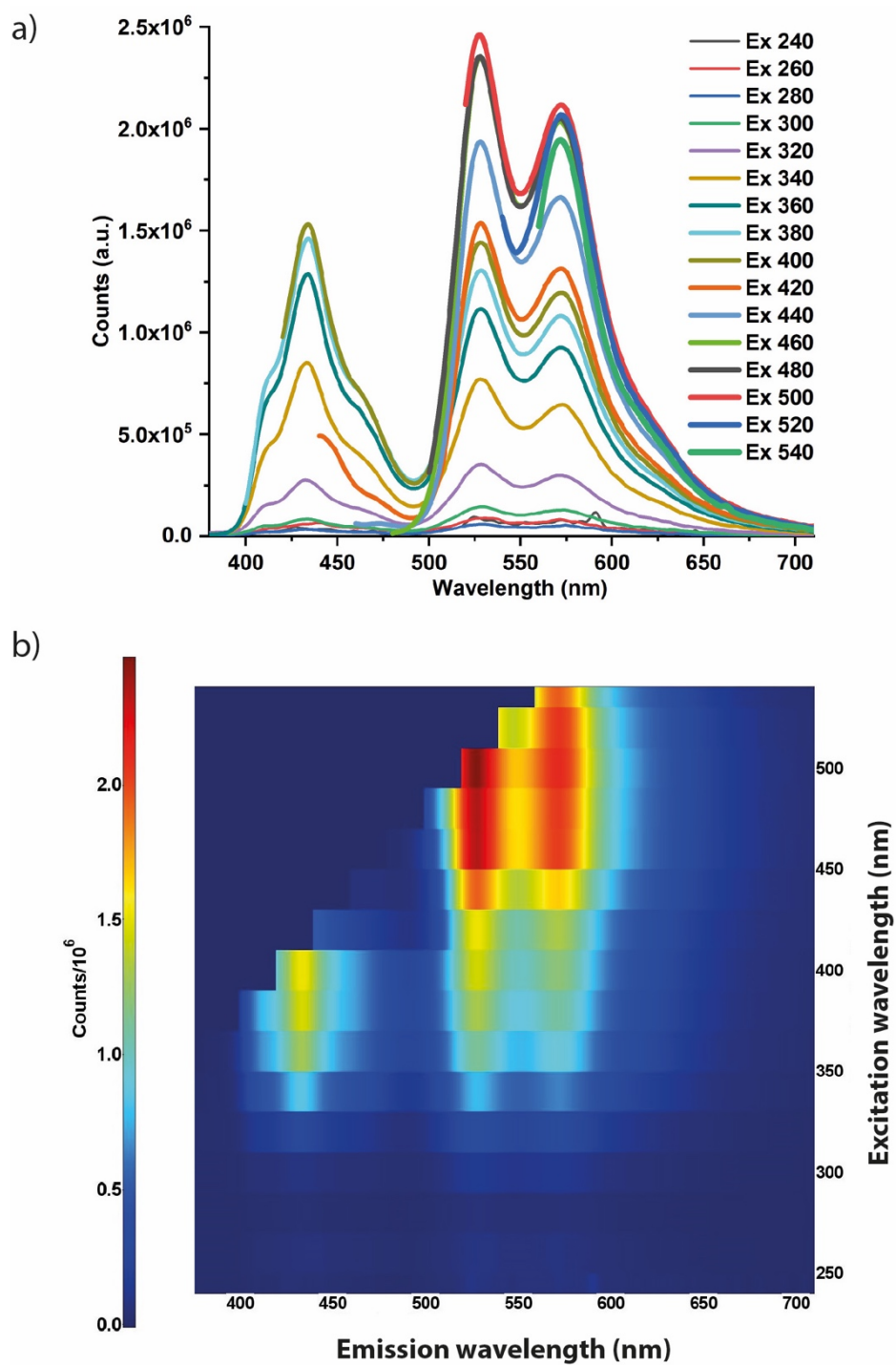

**Figure S4**. (a) Emission spectra of the 3D printed pellet (3 mm thickness) showing the unchanging emission wavelengths subject to different excitation wavelengths, ranging from 240–540 nm. (b) Excitation-emission map corresponding to the series of spectral data in (a). The measurements were performed using the FS5 spectrofluorometer.

## 5 Materials Characterisation

Powder X-ray diffraction (PXRD) of the solid-state powder sample was performed using the Rigaku Miniflex diffractometer. Diffraction data were collected using $2\theta = 5 – 30°$ at a scan rate of 1.0°/min and a step size of 0.01.

Solid-state diffuse reflectance spectra were collected on the Shimadzu UV 2600 equipped with an integration sphere. The sample was prepared by drop coating the powder dispersed in methanol onto a strip of Whatman filter paper. Blank Whatman paper was used to measure the baseline correction and background.

Atomic force microscopy (AFM) imaging on a sample drop-casted onto a 1 $cm^2$ silicon wafer substrate was performed using the neaspec neaSNOM instrument, operating under the tapping mode. Height topography images were collected using the Scout350 probe (NuNano), which has a nominal tip radius of 5 nm, a spring constant of 42 N/m and resonant frequency of 350 kHz.

Thermogravimetric analysis (TGA) was performed on the TA Instrument Q50 instrument, employing at a heating rate of 20 °C/min from 50-650 °C under constant nitrogen flow.

Emission and excitation spectra were measured on the FS5 spectrofluorometer (Edinburgh Instruments) equipped with Xenon lamp light source and a standard detector. Excitation and emission bandwidths were adjusted to maximise the signal for each sample. Fluorescence lifetime decay was studied using the time-correlated single photon counting (TCSPC) technique employing the 363.5nm EPLED picosecond pulsed laser source (Edinburgh Instruments). The quantum yield (QY) measurements of the solid-state and solution samples were performed using the SC-30 integrating sphere module. Lifetime and QY data were analysed using the Fluoracle software. The parameters used in the FS5 spectrofluorometric measurements are listed below.

**Excitation: A=Fluorescein Solution**

    Excitation Scan Parameters:

    Ex Wavelength (nm)      : 250.00 to 570.00 step 1.00

    Dwell Time (s)           : 0.200 per repeat

    Repeats                   : 2

Ex Arm Parameters (Scan Arm):

Bandwidth (nm)            : 1.00

Em Arm Parameters (Fixed Arm):

Wavelength (nm)           : 520.00

Bandwidth (nm)            : 1.00

**Emission: A=Fluorescein Solution**

    Emission Scan Parameters:

    Em Wavelength (nm)        : 460.00 to 750.00 step 1.00

    Dwell Time (s)            : 0.200 per repeat

    Repeats                   : 1

    Ex Arm Parameters (Fixed Arm):

    Wavelength (nm)           : 450.00

    Bandwidth (nm)            : 1.00

    Em Arm Parameters (Scan Arm):

    Bandwidth (nm)            : 1.00

**Excitation: B=Rhodamine B Solution**

    Excitation Scan Parameters:

    Ex Wavelength (nm)        : 250.00 to 650.00 step 0.50

    Dwell Time (s)            : 0.200 per repeat

    Repeats                   : 2

    Ex Arm Parameters (Scan Arm):

    Bandwidth (nm)            : 1.00

    Em Arm Parameters (Fixed Arm):

    Grating                   : Vis

    Wavelength (nm)           : 600.00

    Bandwidth (nm)            : 0.50

**Emission: B=Rhodamine B Solution**

Emission Scan Parameters:

Em Wavelength (nm)        : 480.00 to 750.00 step 1.00

Dwell Time (s)            : 0.200 per repeat

Repeats                   : 1

Ex Arm Parameters (Fixed Arm):

Wavelength (nm)           : 470.00

Bandwidth (nm)            : 1.00

Em Arm Parameters (Scan Arm):

Bandwidth (nm)            : 1.00

**Excitation: A+B Solution**

Excitation Scan Parameters:

Ex Wavelength (nm)        : 250.00 to 650.00 step 1.00

Dwell Time (s)            : 0.200 per repeat

Repeats                   : 2

Ex Arm Parameters (Scan Arm):

Bandwidth (nm)            : 1.00

Em Arm Parameters (Fixed Arm):

Wavelength (nm)           : 550.00

Bandwidth (nm)            : 1.00

**Emission: A+B Solution**

Emission Scan Parameters:

Em Wavelength (nm)        : 460.00 to 750.00 step 1.00

Dwell Time (s)            : 0.200 per repeat

Repeats                   : 1

Ex Arm Parameters (Fixed Arm):

Wavelength (nm)           : 450.00

Bandwidth (nm)            : 1.00

Em Arm Parameters (Scan Arm):

Bandwidth (nm) : 1.00

**Excitation: A+B@ZIF-8 Powder**

    Excitation Scan Parameters:

    Ex Wavelength (nm) : 300.00 to 400.00 step 0.50

    Dwell Time (s) : 0.500 per repeat

    Repeats : 1

    Ex Arm Parameters (Scan Arm):

    Bandwidth (nm) : 2.31

    Em Arm Parameters (Fixed Arm):

    Wavelength (nm) : 620.00

    Bandwidth (nm) : 0.16

**Emission: A+B@ZIF-8 Powder**

    Emission Scan Parameters:

    Em Wavelength (nm) : 480.00 to 750.00 step 1.00

    Dwell Time (s) : 0.100 per repeat

    Repeats : 1

    Ex Arm Parameters (Fixed Arm):

    Wavelength (nm) : 400.00

    Bandwidth (nm) : 4.05

    Em Arm Parameters (Scan Arm):

    Bandwidth (nm) : 0.12

**Excitation: Composite Pellet**

    Excitation Scan Parameters:

    Ex Wavelength (nm) : 240.00 to 605.00 step 0.50

    Dwell Time (s) : 0.500 per repeat

    Repeats : 1

    Ex Arm Parameters (Scan Arm):

Bandwidth (nm)              : 1.50

Em Arm Parameters (Fixed Arm):

Wavelength (nm)             : 610.00

Bandwidth (nm)              : 0.80

**Emission: Composite Pellet**

Emission Scan Parameters:

Em Wavelength (nm)          : 400.00 to 710.00 step 1.00

Dwell Time (s)              : 0.100 per repeat

Repeats                     : 1

Ex Arm Parameters (Fixed Arm):

Wavelength (nm)             : 380.00

Bandwidth (nm)              : 1.00

Em Arm Parameters (Scan Arm):

Bandwidth (nm)              : 0.50

**Excitation: Blank Pellet**

Excitation Scan Parameters:

Ex Wavelength (nm)          : 240.00 to 500.00 step 0.50

Dwell Time (s)              : 0.500 per repeat

Repeats                     : 1

Ex Arm Parameters (Scan Arm):

Bandwidth (nm)              : 1.00

Em Arm Parameters (Fixed Arm):

Wavelength (nm)             : 520.00

Bandwidth (nm)              : 0.50

**Emission: Blank Pellet**

Emission Scan Parameters:

Em Wavelength (nm)          : 400.00 to 700.00 step 1.00

Dwell Time (s) : 0.100 per repeat

Repeats : 1

Ex Arm Parameters (Fixed Arm):

Wavelength (nm) : 380.00

Bandwidth (nm) : 1.00

Em Arm Parameters (Scan Arm):

Bandwidth (nm) : 0.20

## 5.1 Atomic force microscopy (AFM)

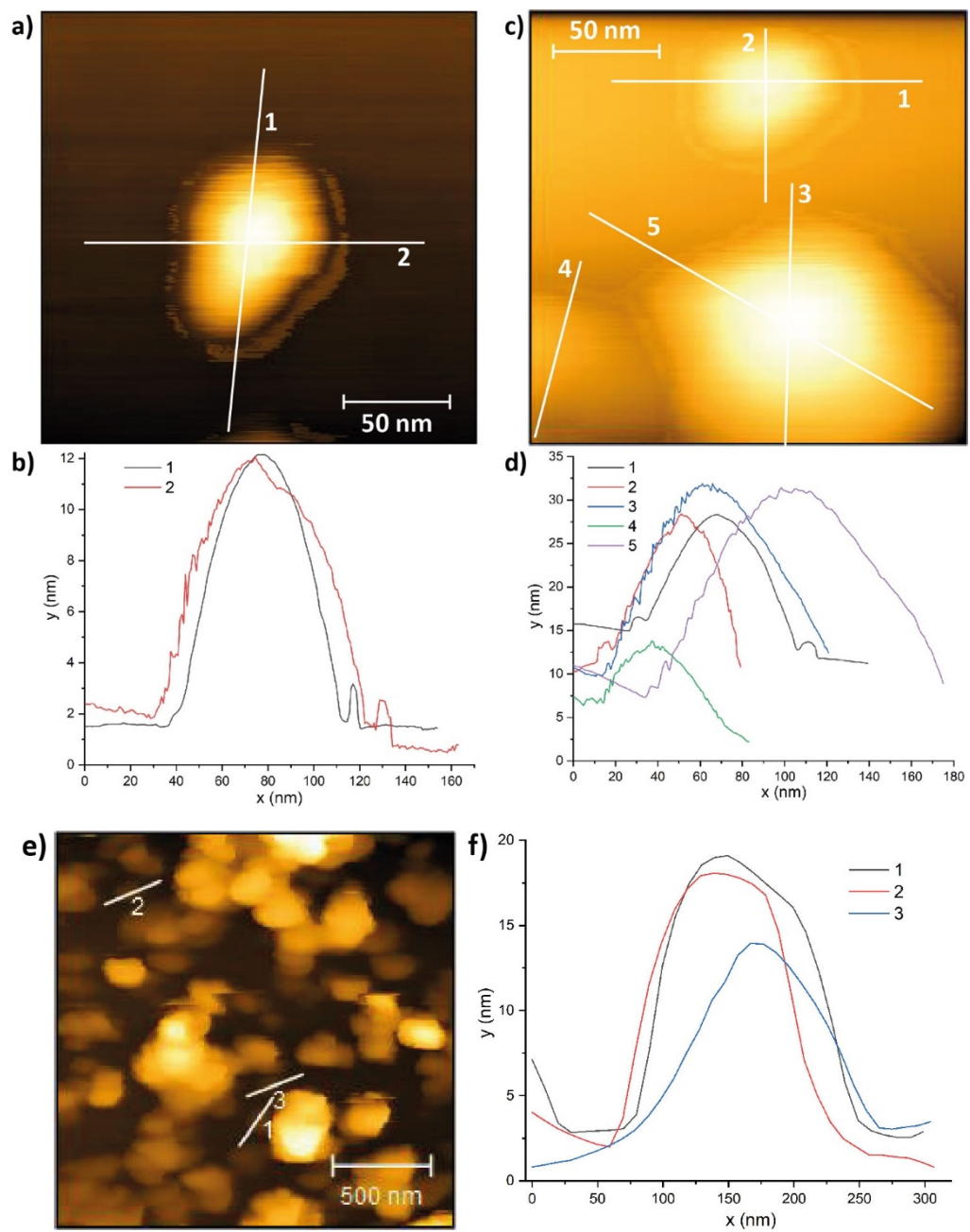

**Figure S5**. Representative AFM images of the nanodiscs of A+B@ZIF-8 (top, a & c), with a graph showing their height profiles (bottom b & d) extracted from the marked regions of the nanodiscs. The nanodiscs exhibit an aspect ratio (width/height) ranging from 5:1 to 10:1

## 5.2 Thermogravimetric analysis (TGA)

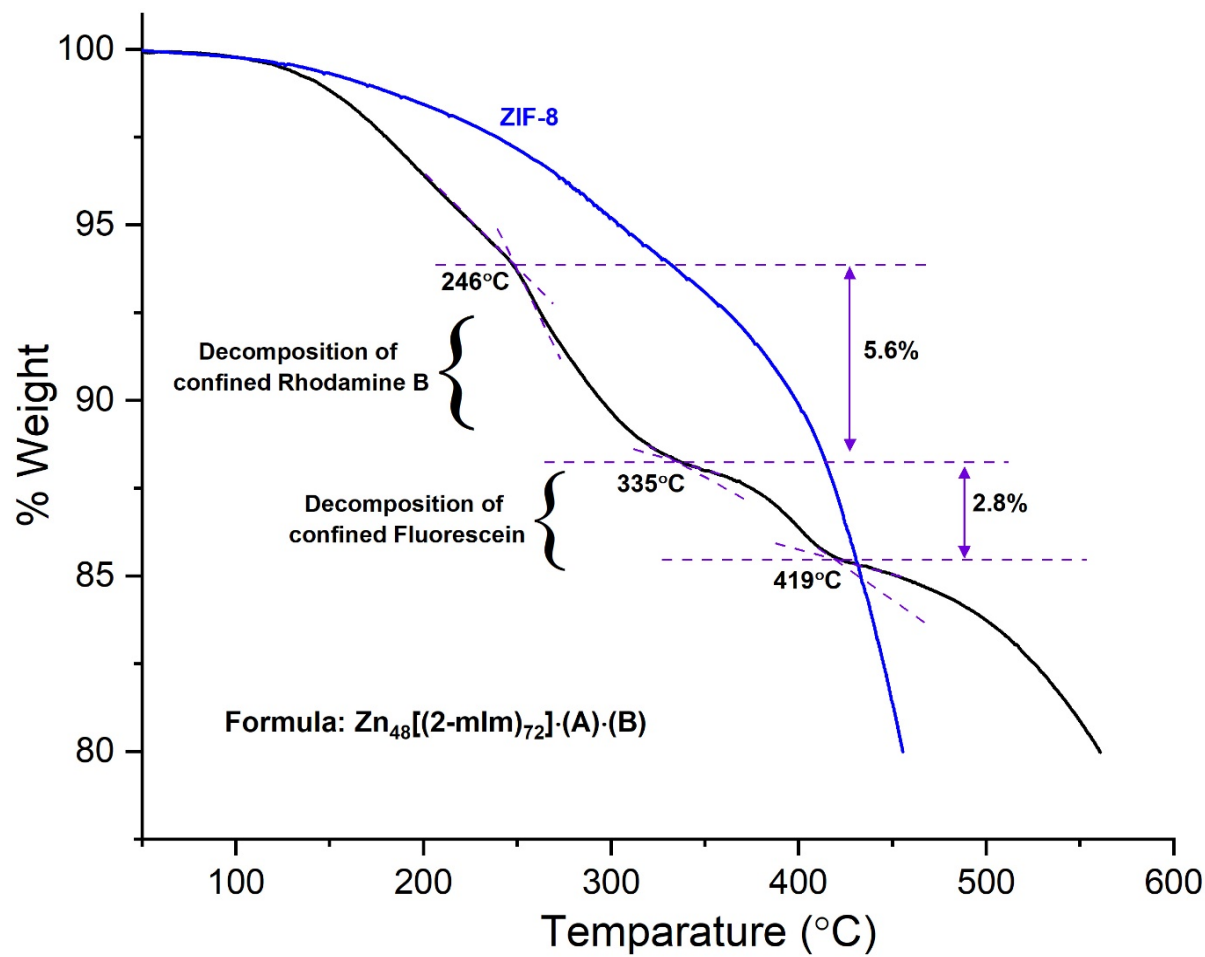

**Figure S6**. TGA of A+B@ZIF-8 powder indicating stepwise degradation of the two confined guest species: Rhodamine B and Fluorescein (A), which were estimated to be at 5.6 wt.% and 2.8 wt.%, respectively. TGA of pristine ZIF-8 is shown for comparison, which was synthesised using the same conditions to that of A+B@ZIF-8 (but without the addition of A+B guest solution).

## 5.3 Diffuse reflectance spectra (DRS)

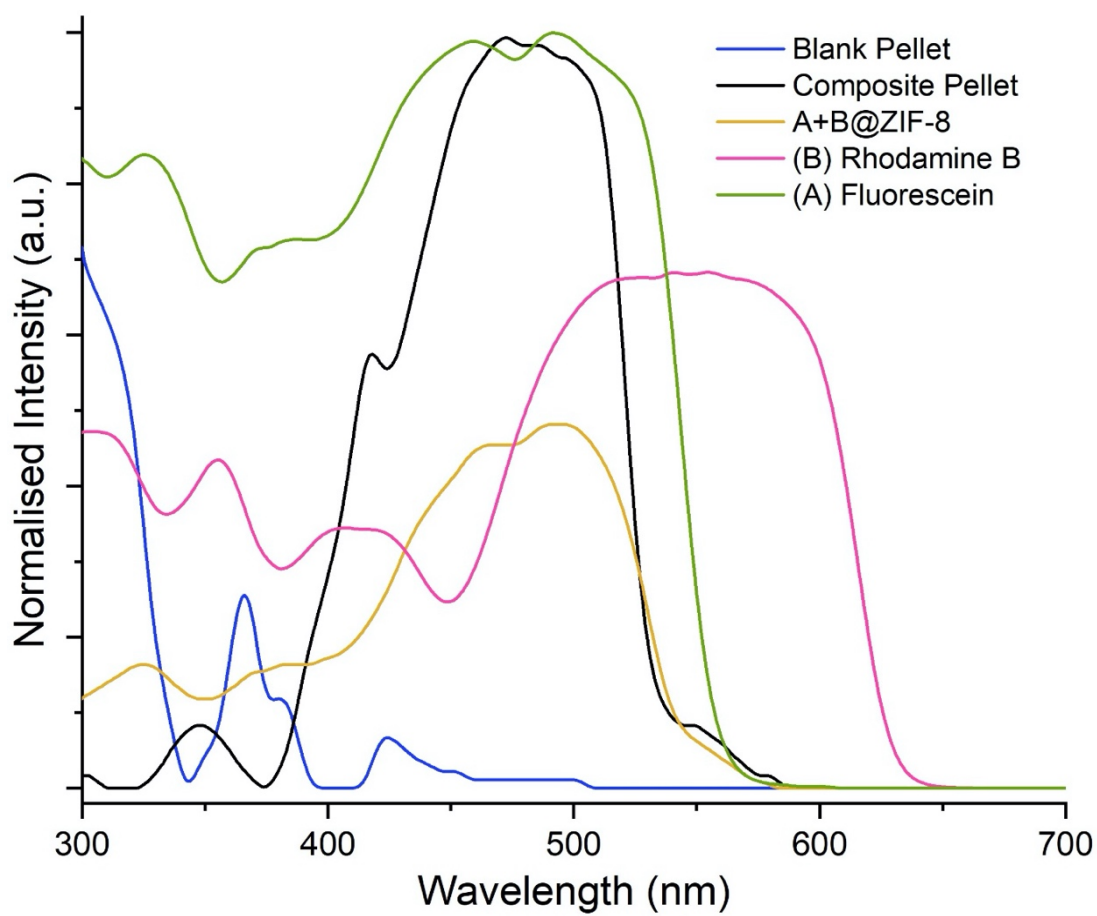

**Figure S7**. Solid-state diffuse reflectance spectra of the 3D printed pellets (blank and composite), powder sample of the guest emitters, and A+B@ZIF-8 powder.

## 5.4 Time-correlated single photon counting (TCSPC)

**Table S1**. Fluorescence lifetime ($\tau$) of different individual materials used in this study. $\alpha$ is the normalised pre-exponential factor.

| Material | $\lambda_{em}$ (nm) | $\alpha$ | $\chi^2$ | $\tau$ (ns) | Rel. % |
|---|---|---|---|---|---|
| (A) Fluorescein Solution | 515 | 0.193 | 1.331 | 4.0859 | 100 |
| (B) Rhodamine B Solution | 571 | 0.215 | 1.315 | 2.4137 | 100 |
| A+B Solution | 515 | 0.194 | 1.184 | 4.0236 | 100 |
| | 571 | 0.197 | 1.184 | 4.0756 | 100 |
| A+B@ZIF-8 Powder | 541 | 0.186 | 1.387 | 1.59 | 52.6 |
| | | 0.077 | | 3.46 | 47.4 |
| | 577 | 0.223 | 1.469 | 4.78 | 100 |
| 'Blank' 3D printed pellet (Formlab's clear resin) | 440 | 0.182 | 1.345 | 15.4 | 100 |
| 'Composite' 3D printed pellet (A+B@ZIF-8 combined with clear resin) | 433 | 0.303 | 1.319 | 1.2816 | 88.96 |
| | | 0.004 | | 11.711 | 11.04 |
| | 528 | -0.141 | 1.171 | 1.0377 | 10.13 |
| | | 0.288 | | 3.8009 | 75.97 |
| | | 0.015 | | 13.9975 | 14.57 |
| | 571 | -0.382 | 1.067 | 2.15 | 24 |
| | | 0.46 | | 4.97 | 66 |
| | | 0.02 | | 15.58 | 9.1 |

## 5.5 Quantum yield (QY)

**Table S2**. Quantum yield (%) of different individual materials used in this study.

| Material / Species | Quantum Yield (%) |
|---|---|
| A = Fluorescein | 40.28 |
| B = Rhodamine B | 35.53 |
| A+B Solution | 68.01 |
| A+B@ZIF-8 Powder | 47.31 |
| Composite Pellet (3D printed) | 43.56 |
| Blank Pellet (3D printed) | 62.32 |

**Quantum Yield: A=Fluorescein Solution**

 Emission Scan Parameters:

 Em Wavelength (nm)  : 450.00 to 750.00 step 1.00

 Dwell Time (s)  : 0.200 per repeat

 Repeats  : 1

 Ex Arm Parameters (Fixed Arm):

 Wavelength (nm)  : 460.00

 Bandwidth (nm)  : 3.50

 Mono Type  : FS5

 Em Arm Parameters (Scan Arm):

 Bandwidth (nm)  : 0.35

**Quantum Yield: B=Rhodamine B Solution**

 Emission Scan Parameters:

 Em Wavelength (nm)  : 460.00 to 750.00 step 1.00

 Dwell Time (s)  : 0.200 per repeat

 Repeats  : 1

 Ex Arm Parameters (Fixed Arm):

Wavelength (nm)            : 470.00

Bandwidth (nm)             : 3.50

Em Arm Parameters (Scan Arm):

Bandwidth (nm)             : 0.35

**Quantum Yield: A+B Solution**

Emission Scan Parameters:

Em Wavelength (nm)         : 450.00 to 750.00 step 1.00

Dwell Time (s)             : 0.200 per repeat

Repeats                    : 1

Ex Arm Parameters (Fixed Arm):

Wavelength (nm)            : 460.00

Bandwidth (nm)             : 3.50

Em Arm Parameters (Scan Arm):

Bandwidth (nm)             : 0.35

**Quantum Yield: A+B@ZIF-8 Powder**

Emission Scan Parameters:

Em Wavelength (nm)         : 460.00 to 750.00 step 1.00

Dwell Time (s)             : 0.200 per repeat

Repeats                    : 1

Ex Arm Parameters (Fixed Arm):

Wavelength (nm)            : 480.00

Bandwidth (nm)             : 4.00

Em Arm Parameters (Scan Arm):

Bandwidth (nm)             : 0.40

**Quantum Yield: Composite Pellet**

Emission Scan Parameters:

Em Wavelength (nm)         : 350.00 to 750.00 step 1.00

Dwell Time (s)              : 0.200 per repeat

Repeats                     : 1

Ex Arm Parameters (Fixed Arm):

Wavelength (nm)             : 370.00

Bandwidth (nm)              : 9.00

Em Arm Parameters (Scan Arm):

Bandwidth (nm)              : 0.80

**Quantum Yield: Blank Pellet**

Emission Scan Parameters:

Em Wavelength (nm)          : 370.00 to 650.00 step 1.00

Dwell Time (s)              : 0.200 per repeat

Repeats                     : 1

Ex Arm Parameters (Fixed Arm):

Wavelength (nm)             : 390.00

Bandwidth (nm)              : 4.00

Em Arm Parameters (Scan Arm):

Bandwidth (nm)              : 0.40

**Table S3**. Change in the quantum yield (QY) of the 3D printed disc pellet upon exposure to water, UV, and heat. Note that these are accelerated ageing conditions where more extreme test conditions were applied compared to a typical LED.

|  | QY (%) | | |
|---|---|---|---|
| **Exposure time** | 0 hr | 1 hr | 24 hr |
| **Water Immersion** | 43.56 | 42.42 | 40.34 |
| **UV radiation at 400 nm** | 43.56 | 41.57 | 38.63 |
| **Heating at 60 °C** | 43.56 | 42.19 | 39.18 |

**Table S4**. Characterisation of the lighting performance of the 3D printed composite pellets when irradiated under the 400-nm UV LED, measured using the UPRtek photospectrometer. Note: LUX = spectral illuminance; CRI = colour rendering index; CCT = chromaticity colour temperature (in K).

| Thickness of 3D printed composite pellet (mm) → | 0.5 | 1.5 | 2.2 | 3.0 | 4.4 | 5.2 |
|---|---|---|---|---|---|---|
| **LUX** | 1376.906 | 1824.283 | 1419.112 | 1332.541 | 922.5844 | 808.158 |
| **CRI** | - | 46.64893 | 43.39875 | 42.12814 | 42.53244 | 43.5467 |
| **CCT (K)** | - | 8325 | 6095 | 4701 | 4180 | 3701 |

## 5.6 Lifetime measurements using TCSPC

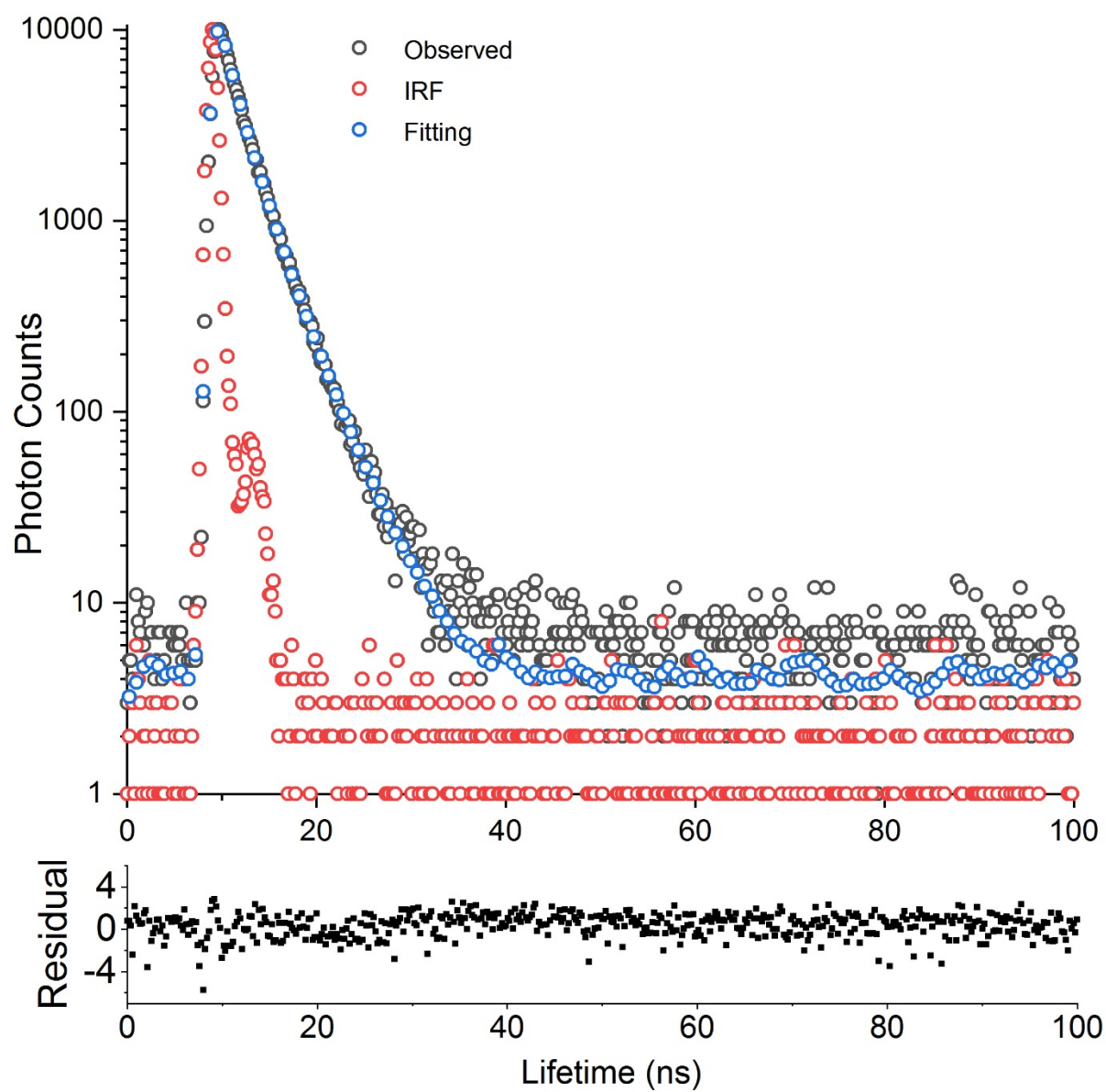

**Figure S8**. Lifetime decay profile of A+B@ZIF-8 powder measured for the emission maximum at 541 nm. IRF is the instrument response function.

**Decay: A+B@ZIF-8 Powder@541nm emission**

    Lifetime Parameters:

    Time Range (ns)                : 0.00000 to 199.80469 step 0.19531

    Channel Range                 : 0 to 1023

    Mode                            : TCSPC

    TAC (ns)                       : 200

    Delay (ns)                      : 0

| | |
|---|---|
| Time Calibration (ns) | : 0.19531 |
| Reps | : 1 |
| Acq Time (s) | : 20.3 |
| Ex Arm Parameters: | |
| WaveLength (nm) | : 362.50 |
| Bandwidth (nm) | : 0.01 |
| Lightpath | : TCSPC Diode |
| Em Arm Parameters: | |
| WaveLength (nm) | : 541.00 |
| Bandwidth (nm) | : 5.70 |

**IRF: A+B@ZIF-8 Powder@541nm emission**

| | |
|---|---|
| Lifetime Parameters: | |
| Time Range (ns) | : 0.00000 to 199.80469 step 0.19531 |
| Channel Range | : 0 to 1023 |
| Mode | : TCSPC |
| TAC (ns) | : 200 |
| Delay (ns) | : 0 |
| Time Calibration (ns) | : 0.19531 |
| Reps | : 1 |
| Acq Time (s) | : 7.1 |
| Ex Arm Parameters: | |
| Intensity | : 50 |
| WaveLength (nm) | : 362.50 |
| Bandwidth (nm) | : 0.01 |
| Lightpath | : TCSPC Diode |
| Em Arm Parameters: | |
| WaveLength (nm) | : 362.50 |
| Bandwidth (nm) | : 5.70 |

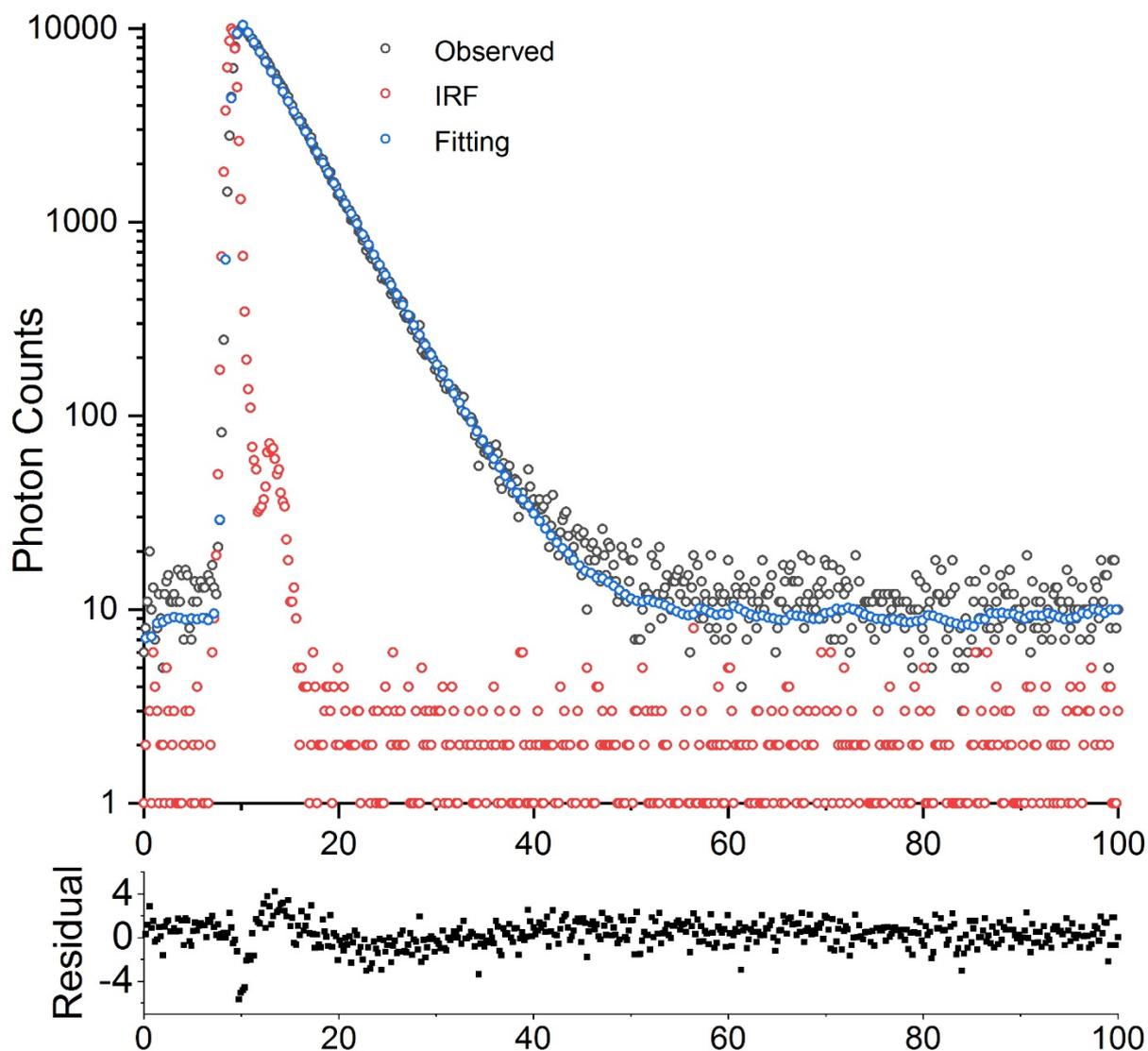

**Figure S9**. Lifetime decay profile of A+B@ZIF-8 powder measured for the emission maximum at 577 nm.

**Decay: A+B@ZIF-8 Powder@577nm emission**

    Lifetime Parameters:

| | |
|---|---|
| Time Range (ns) | : 0.00000 to 199.80469 step 0.19531 |
| Channel Range | : 0 to 1023 |
| Mode | : TCSPC |
| TAC (ns) | : 200 |
| Delay (ns) | : 0 |
| Time Calibration (ns) | : 0.19531 |
| Reps | : 1 |
| Acq Time (s) | : 39.2 |
| Ex Arm Parameters: | |
| WaveLength (nm) | : 362.50 |
| Bandwidth (nm) | : 0.01 |
| Lightpath | : TCSPC Diode |
| Em Arm Parameters: | |
| WaveLength (nm) | : 577.00 |
| Bandwidth (nm) | : 5.70 |

**IRF: A+B@ZIF-8 Powder@577nm emission**

| | |
|---|---|
| Lifetime Parameters: | |
| Time Range (ns) | : 0.00000 to 199.80469 step 0.19531 |
| Channel Range | : 0 to 1023 |
| Mode | : TCSPC |
| TAC (ns) | : 200 |
| Delay (ns) | : 0 |
| Time Calibration (ns) | : 0.19531 |
| Reps | : 1 |
| Acq Time (s) | : 7.1 |
| Ex Arm Parameters: | |
| WaveLength (nm) | : 362.50 |
| Bandwidth (nm) | : 0.01 |
| Lightpath | : TCSPC Diode |
| Em Arm Parameters: | |
| WaveLength (nm) | : 362.50 |
| Bandwidth (nm) | : 5.70 |

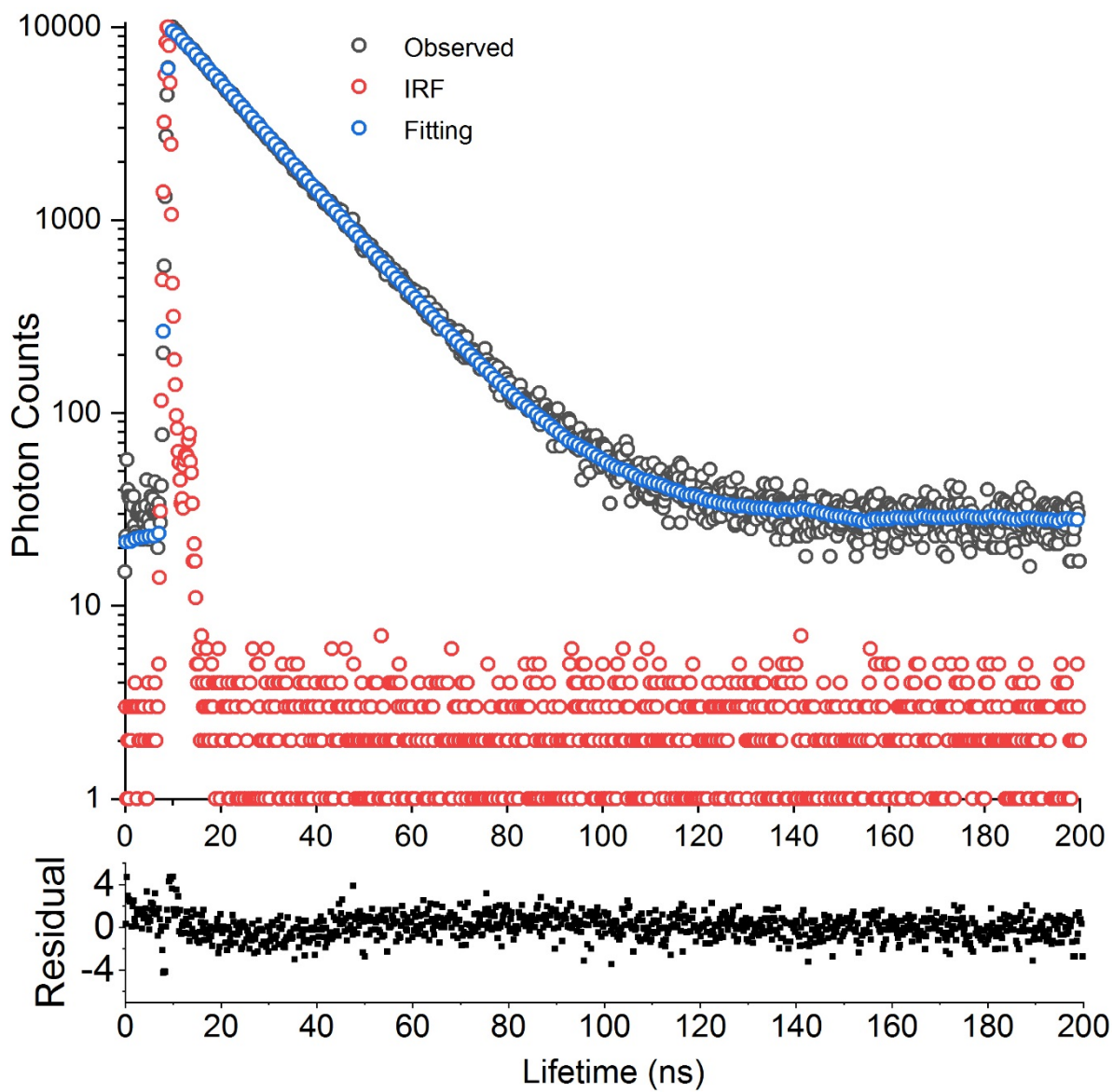

**Figure S10**. Lifetime decay profile of blank pellet (3D printed pure resin) measured for the emission maximum at 440 nm.

**Decay: Blank Pellet@440nm emission**

    Lifetime Parameters:

| | |
|---|---|
| Time Range (ns) | : 0.00000 to 199.80469 step 0.19531 |
| Channel Range | : 0 to 1023 |
| Mode | : TCSPC |
| TAC (ns) | : 200 |
| Delay (ns) | : 0 |
| Time Calibration (ns) | : 0.19531 |
| Reps | : 1 |
| Acq Time (m:s) | : 01:38.3 |
| Ex Arm Parameters: | |
| WaveLength (nm) | : 362.50 |
| Bandwidth (nm) | : 0.01 |
| Lightpath | : TCSPC Diode |
| Em Arm Parameters: | |
| WaveLength (nm) | : 440.00 |
| Bandwidth (nm) | : 1.65 |

**IRF: Blank Pellet@440nm emission**

| | |
|---|---|
| Lifetime Parameters: | |
| Time Range (ns) | : 0.00000 to 199.80469 step 0.19531 |
| Channel Range | : 0 to 1023 |
| Mode | : TCSPC |
| TAC (ns) | : 200 |
| Delay (ns) | : 0 |
| Time Calibration (ns) | : 0.19531 |
| Reps | : 1 |
| Acq Time (s) | : 7.7 |
| Ex Arm Parameters: | |
| WaveLength (nm) | : 362.50 |
| Bandwidth (nm) | : 0.01 |
| Lightpath | : TCSPC Diode |
| Em Arm Parameters: | |
| WaveLength (nm) | : 362.50 |
| Bandwidth (nm) | : 1.49 |

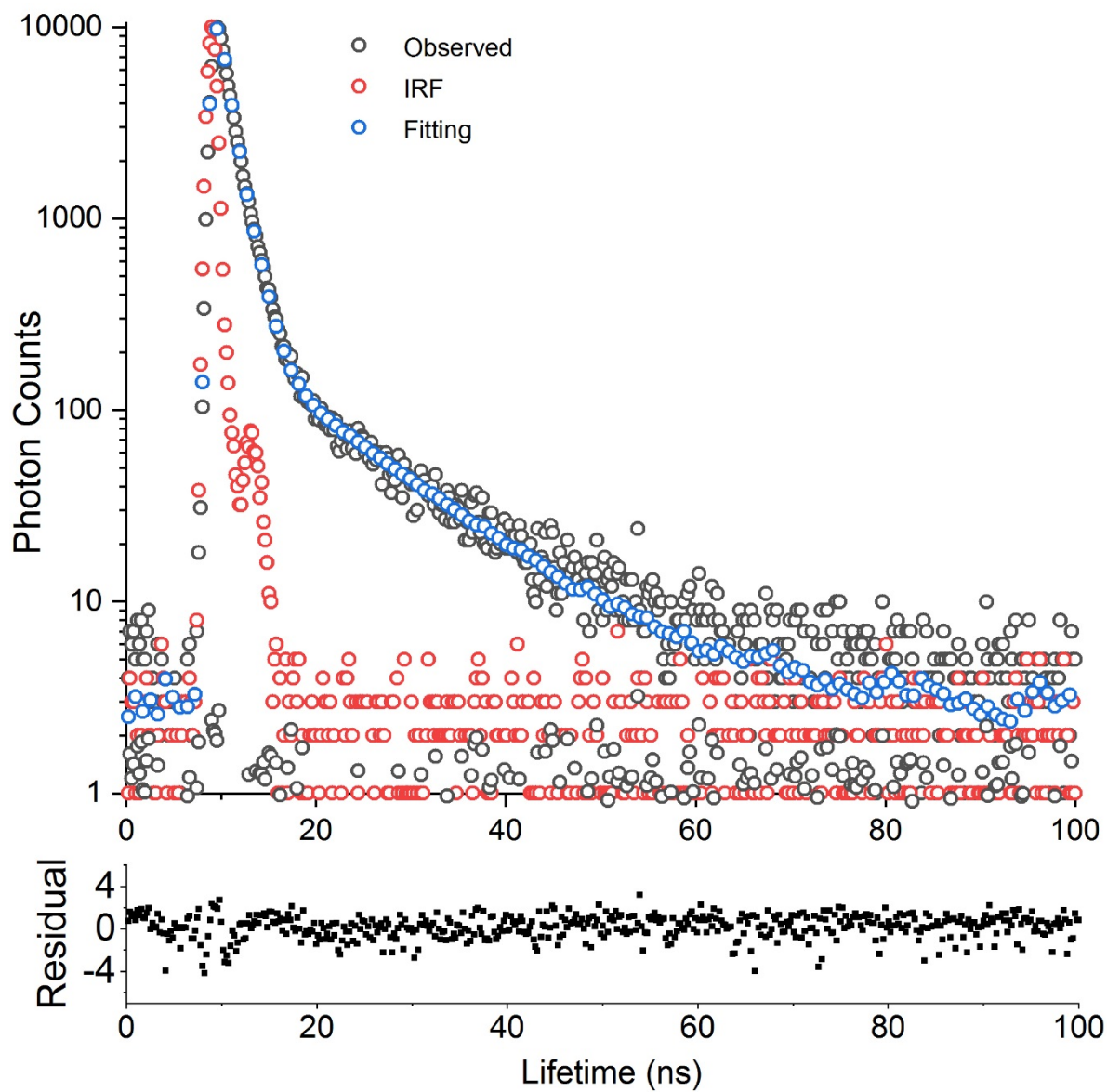

**Figure S11.** Lifetime decay profile of composite 3D printed pellet measured for the emission maximum at 433 nm.

**Decay: Composite Pellet@433nm emission**

    Lifetime Parameters:

| | |
|---|---|
| Time Range (ns) | : 0.00000 to 199.80469 step 0.19531 |
| Channel Range | : 0 to 1023 |
| Mode | : TCSPC |
| TAC (ns) | : 200 |
| Delay (ns) | : 0 |
| Time Calibration (ns) | : 0.19531 |
| Reps | : 1 |
| Acq Time (s) | : 7.8 |
| Ex Arm Parameters: | |
| WaveLength (nm) | : 362.50 |
| Bandwidth (nm) | : 0.01 |
| Lightpath | : TCSPC Diode |
| Em Arm Parameters: | |
| WaveLength (nm) | : 433.00 |
| Bandwidth (nm) | : 3.33 |

**IRF: Composite Pellet@433nm emission**

| | |
|---|---|
| Lifetime Parameters: | |
| Time Range (ns) | : 0.00000 to 199.80469 step 0.19531 |
| Channel Range | : 0 to 1023 |
| Mode | : TCSPC |
| TAC (ns) | : 200 |
| Delay (ns) | : 0 |
| Time Calibration (ns) | : 0.19531 |
| Reps | : 1 |
| Acq Time (s) | : 3.10 |
| Ex Arm Parameters: | |
| WaveLength (nm) | : 362.50 |
| Bandwidth (nm) | : 0.01 |
| Lightpath | : TCSPC Diode |
| Em Arm Parameters: | |
| WaveLength (nm) | : 362.50 |
| Bandwidth (nm) | : 3.33 |

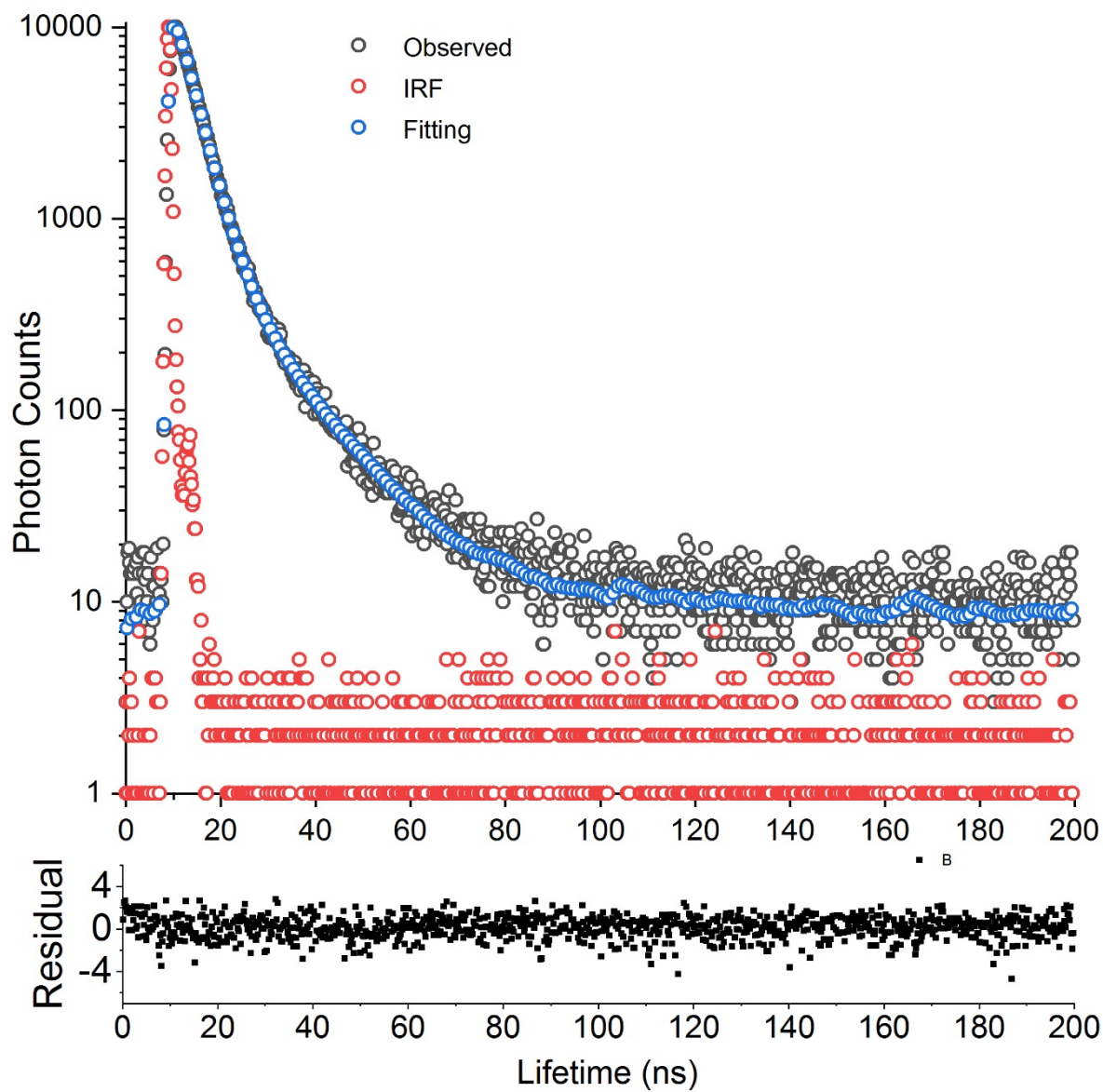

**Figure S12**. Lifetime decay profile of composite 3D printed pellet measured for the emission maximum at 528 nm.

**Decay: Composite Pellet@528nm emission**

    Lifetime Parameters:

| | |
|---|---|
| Time Range (ns) | : 0.00000 to 199.80469 step 0.19531 |
| Channel Range | : 0 to 1023 |
| Mode | : TCSPC |
| TAC (ns) | : 200 |
| Delay (ns) | : 0 |
| Time Calibration (ns) | : 0.19531 |
| Reps | : 1 |
| Acq Time (s) | : 30.0 |
| Ex Arm Parameters: | |
| WaveLength (nm) | : 362.50 |
| Bandwidth (nm) | : 0.01 |
| Lightpath | : TCSPC Diode |
| Em Arm Parameters: | |
| WaveLength (nm) | : 528.00 |
| Bandwidth (nm) | : 3.33 |

**IRF: Composite Pellet@528nm emission**

| | |
|---|---|
| Lifetime Parameters: | |
| Time Range (ns) | : 0.00000 to 199.80469 step 0.19531 |
| Channel Range | : 0 to 1023 |
| Mode | : TCSPC |
| TAC (ns) | : 200 |
| Delay (ns) | : 0 |
| Time Calibration (ns) | : 0.19531 |
| Reps | : 1 |
| Acq Time (s) | : 5.0 |
| Ex Arm Parameters: | |
| WaveLength (nm) | : 362.50 |
| Bandwidth (nm) | : 0.01 |
| Lightpath | : TCSPC Diode |
| Em Arm Parameters: | |
| WaveLength (nm) | : 362.50 |
| Bandwidth (nm) | : 3.33 |

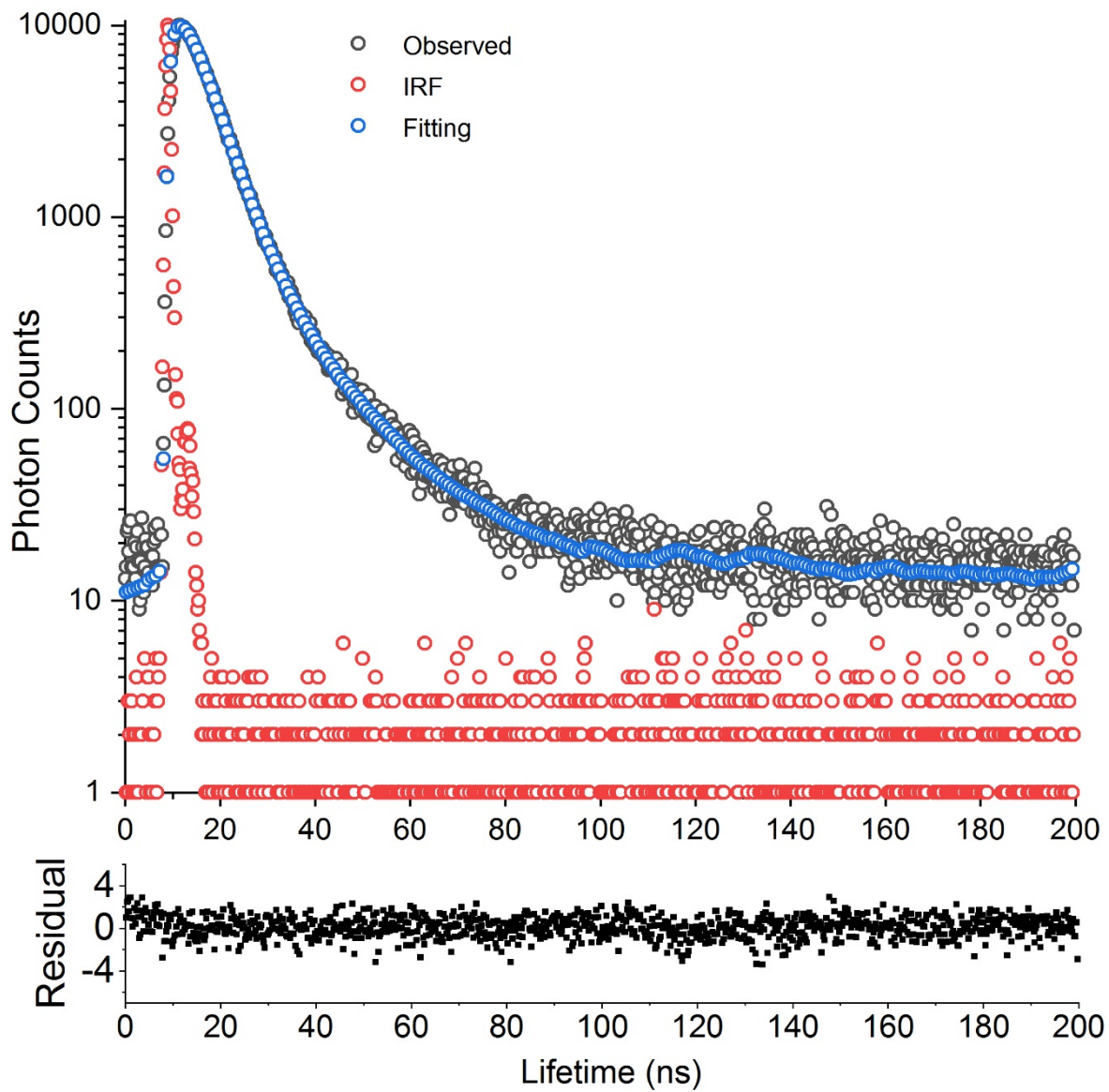

**Figure S13**. Lifetime decay profile of composite 3D printed pellet measured for the emission maximum at 572 nm.

**Decay: Composite Pellet@572nm emission**

    Lifetime Parameters:

| | |
|---|---|
| Time Range (ns) | : 0.00000 to 199.80469 step 0.19531 |
| Channel Range | : 0 to 1023 |
| Mode | : TCSPC |
| TAC (ns) | : 200 |
| Delay (ns) | : 0 |
| Time Calibration (ns) | : 0.19531 |
| Reps | : 1 |
| Acq Time (m:s) | : 01:18.6 |
| Ex Arm Parameters: | |
| WaveLength (nm) | : 362.50 |
| Bandwidth (nm) | : 0.01 |
| Lightpath | : TCSPC Diode |
| Em Arm Parameters: | |
| WaveLength (nm) | : 572.00 |
| Bandwidth (nm) | : 3.33 |

**IRF: Composite Pellet@572nm emission**

| | |
|---|---|
| Lifetime Parameters: | |
| Time Range (ns) | : 0.00000 to 199.80469 step 0.19531 |
| Channel Range | : 0 to 1023 |
| Mode | : TCSPC |
| TAC (ns) | : 200 |
| Delay (ns) | : 0 |
| Time Calibration (ns) | : 0.19531 |
| Reps | : 1 |
| Acq Time (s) | : 8.1 |
| Ex Arm Parameters: | |
| WaveLength (nm) | : 362.50 |
| Bandwidth (nm) | : 0.01 |
| Mono Type | : FS5 |
| Lightpath | : TCSPC Diode |
| Em Arm Parameters: | |
| WaveLength (nm) | : 362.50 |
| Bandwidth (nm) | : 3.33 |

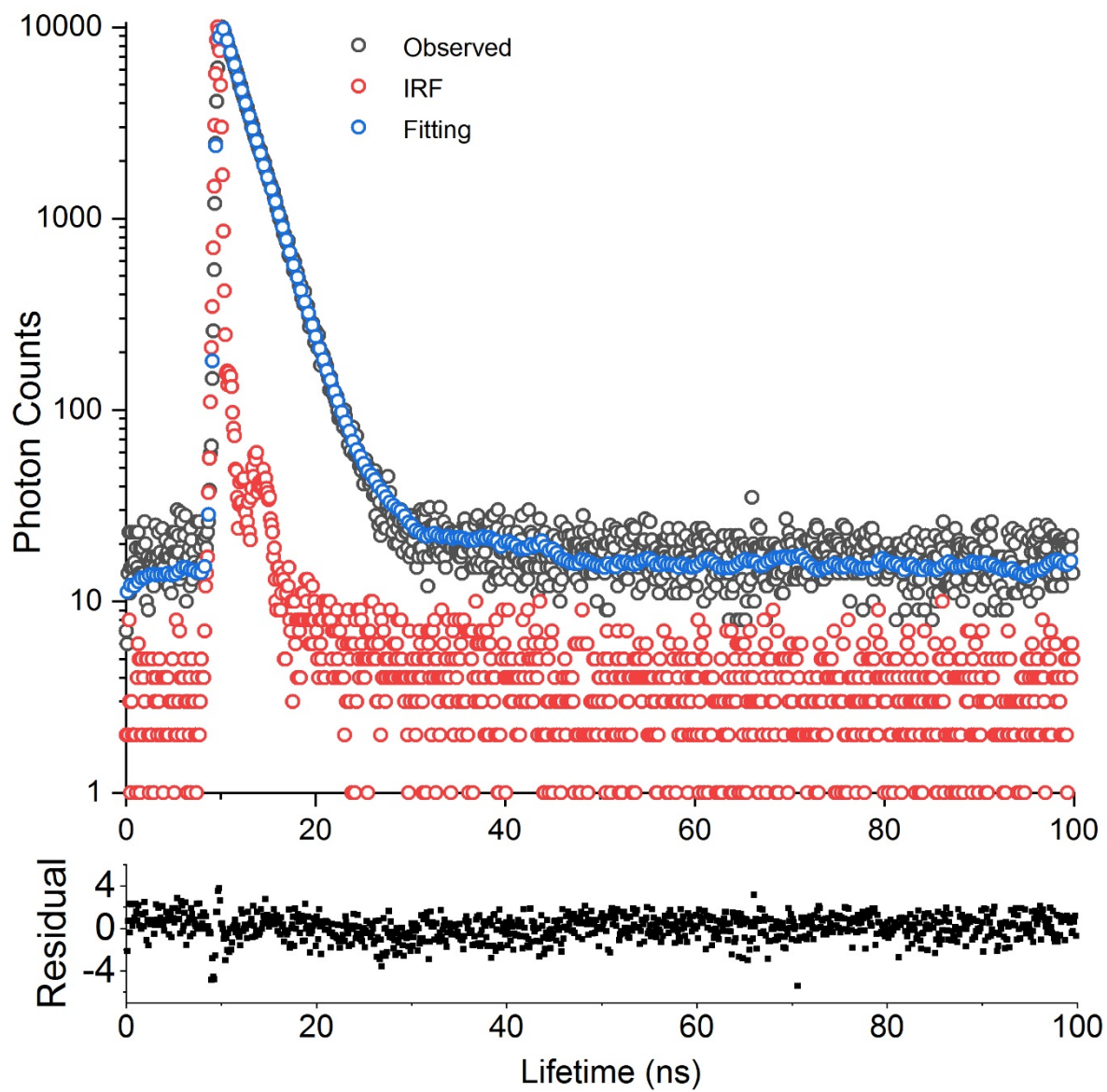

**Figure S14**. Lifetime decay profile of Rhodamine B solution measured for the emission maximum at 571 nm.

**Decay: B=Rhodamine B Solution**

    Lifetime Parameters:

| | |
|---|---|
| Time Range (ns) | : 0.00000 to 99.90234 step 0.09766 |
| Channel Range | : 0 to 1023 |
| Mode | : TCSPC |
| TAC (ns) | : 100 |
| Delay (ns) | : 0 |
| Time Calibration (ns) | : 0.09766 |
| Reps | : 1 |
| Acq Time (m:s) | : 01:12.1 |
| Ex Arm Parameters: | |
| WaveLength (nm) | : 362.50 |
| Bandwidth (nm) | : 0.01 |
| Lightpath | : TCSPC Diode |
| Em Arm Parameters: | |
| WaveLength (nm) | : 571.00 |
| Bandwidth (nm) | : 1.00 |
| Lightpath | : Standard Detector |

**IRF: B=Rhodamine B Solution**

| | |
|---|---|
| Lifetime Parameters: | |
| Time Range (ns) | : 0.00000 to 99.90234 step 0.09766 |
| Channel Range | : 0 to 1023 |
| Mode | : TCSPC |
| TAC (ns) | : 100 |
| Delay (ns) | : 0 |
| Time Calibration (ns) | : 0.09766 |
| Reps | : 1 |
| Acq Time (s) | : 15.7 |
| Ex Arm Parameters: | |
| WaveLength (nm) | : 362.50 |
| Bandwidth (nm) | : 0.01 |
| Lightpath | : TCSPC Diode |
| Em Arm Parameters: | |
| WaveLength (nm) | : 362.50 |
| Bandwidth (nm) | : 4.01 |
| Lightpath | : Standard Detector |

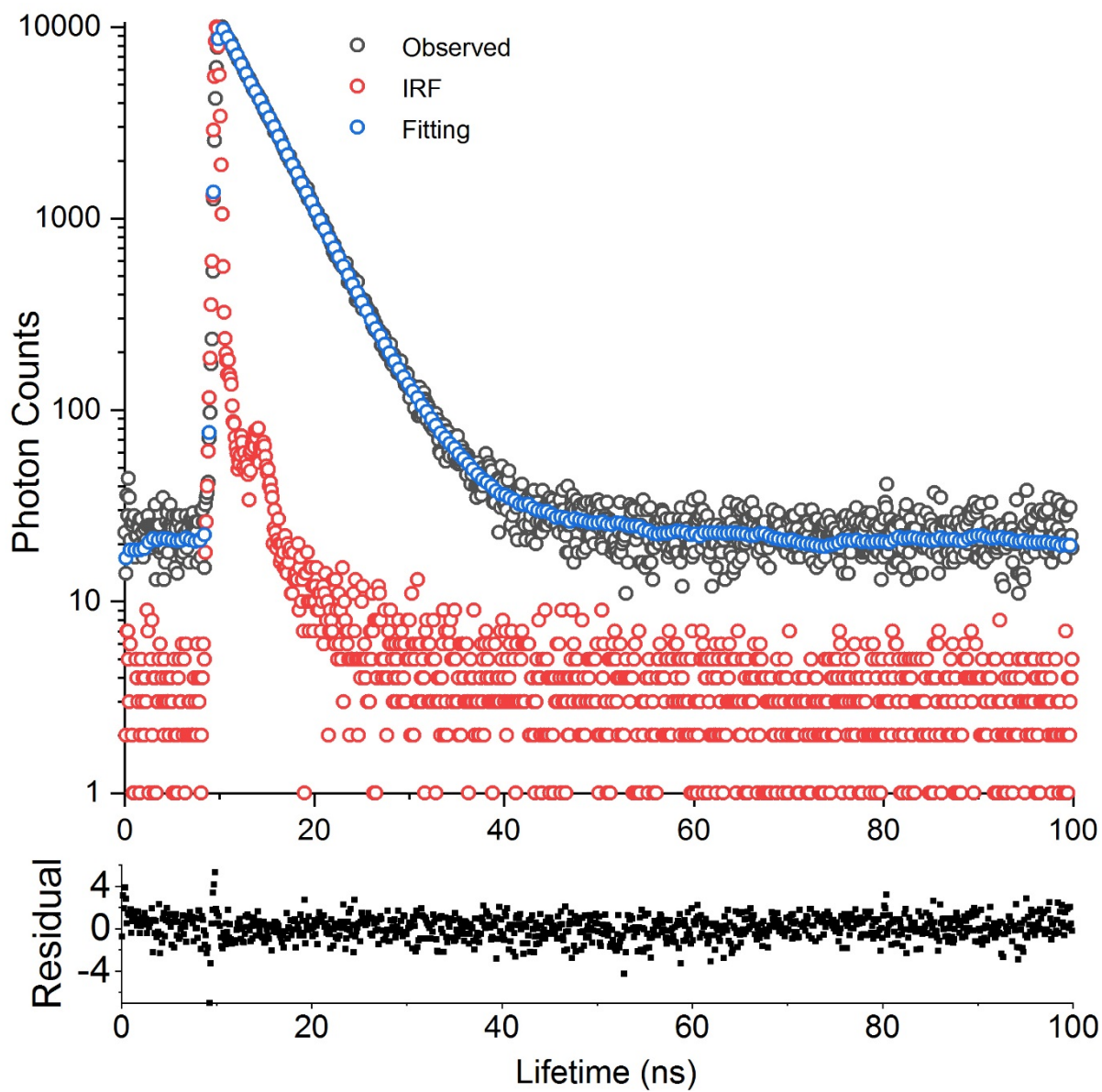

**Figure S15**. Lifetime decay profile of A = Fluorescein solution measured for the emission maximum at 515 nm.

**Decay: A=Fluorescein Solution**

    Lifetime Parameters:

| | |
|---|---|
| Time Range (ns) | : 0.00000 to 99.90234 step 0.09766 |
| Channel Range | : 0 to 1023 |
| Mode | : TCSPC |
| TAC (ns) | : 100 |
| Delay (ns) | : 0 |
| Time Calibration (ns) | : 0.09766 |
| Reps | : 1 |
| Acq Time (m:s) | : 01:33.8 |
| Ex Arm Parameters: | |
| WaveLength (nm) | : 362.50 |
| Bandwidth (nm) | : 0.01 |
| Lightpath | : TCSPC Diode |
| Em Arm Parameters: | |
| WaveLength (nm) | : 515.00 |
| Bandwidth (nm) | : 1.00 |

**IRF: A=Fluorescein Solution**

| | |
|---|---|
| Lifetime Parameters: | |
| Time Range (ns) | : 0.00000 to 99.90234 step 0.09766 |
| Channel Range | : 0 to 1023 |
| Mode | : TCSPC |
| TAC (ns) | : 100 |
| Time Calibration (ns) | : 0.09766 |
| Reps | : 1 |
| Acq Time (s) | : 11.4 |
| Ex Arm Parameters: | |
| Intensity | : 50 |
| WaveLength (nm) | : 362.50 |
| Bandwidth (nm) | : 0.01 |
| Lightpath | : TCSPC Diode |
| Em Arm Parameters: | |
| WaveLength (nm) | : 362.50 |
| Bandwidth (nm) | : 4.19 |

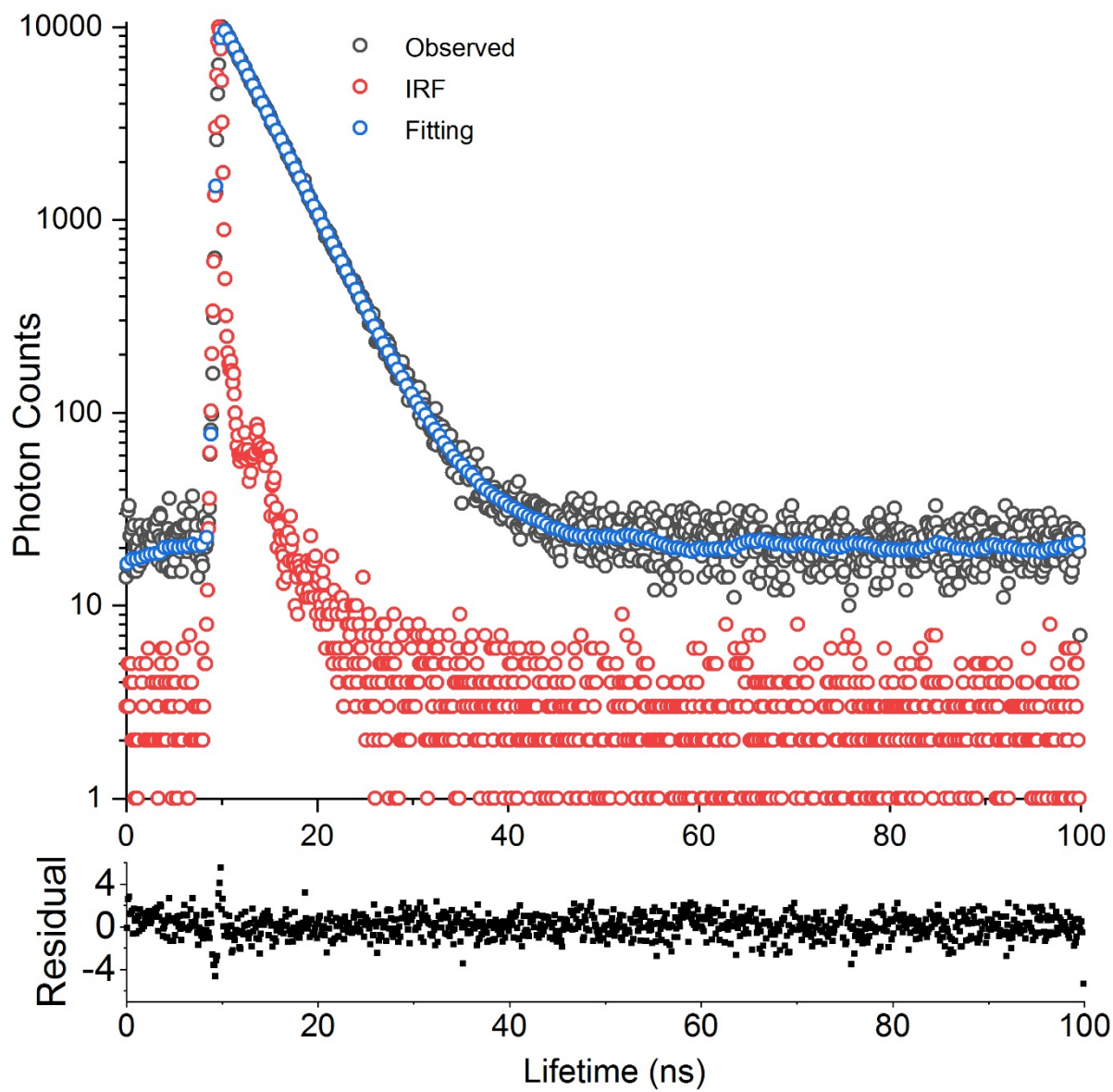

**Figure S16**. Lifetime decay profile of A+B solution (i.e. Rhodamine B and Fluorescein) measured for the emission maximum at 515 nm.

**Decay: A+B Solution@515nm emission**

    Lifetime Parameters:

| | |
|---|---|
| Time Range (ns) | : 0.00000 to 99.90234 step 0.09766 |
| Channel Range | : 0 to 1023 |
| Mode | : TCSPC |
| TAC (ns) | : 100 |
| Delay (ns) | : 0 |
| Time Calibration (ns) | : 0.09766 |
| Reps | : 1 |
| Acq Time (m:s) | : 01:30.7 |
| Ex Arm Parameters: | |
| WaveLength (nm) | : 362.50 |
| Bandwidth (nm) | : 0.01 |
| Lightpath | : TCSPC Diode |
| Em Arm Parameters: | |
| WaveLength (nm) | : 515.00 |
| Bandwidth (nm) | : 1.00 |
| Lightpath | : Standard Detector |

**IRF: A+B Solution @515nm emission**

| | |
|---|---|
| Lifetime Parameters: | |
| Time Range (ns) | : 0.00000 to 99.90234 step 0.09766 |
| Channel Range | : 0 to 1023 |
| Mode | : TCSPC |
| TAC (ns) | : 100 |
| Delay (ns) | : 0 |
| Time Calibration (ns) | : 0.09766 |
| Reps | : 1 |
| Acq Time (s) | : 11.3 |
| Ex Arm Parameters: | |
| WaveLength (nm) | : 362.50 |
| Bandwidth (nm) | : 0.01 |
| Lightpath | : TCSPC Diode |
| Em Arm Parameters: | |
| WaveLength (nm) | : 362.50 |
| Bandwidth (nm) | : 4.42 |

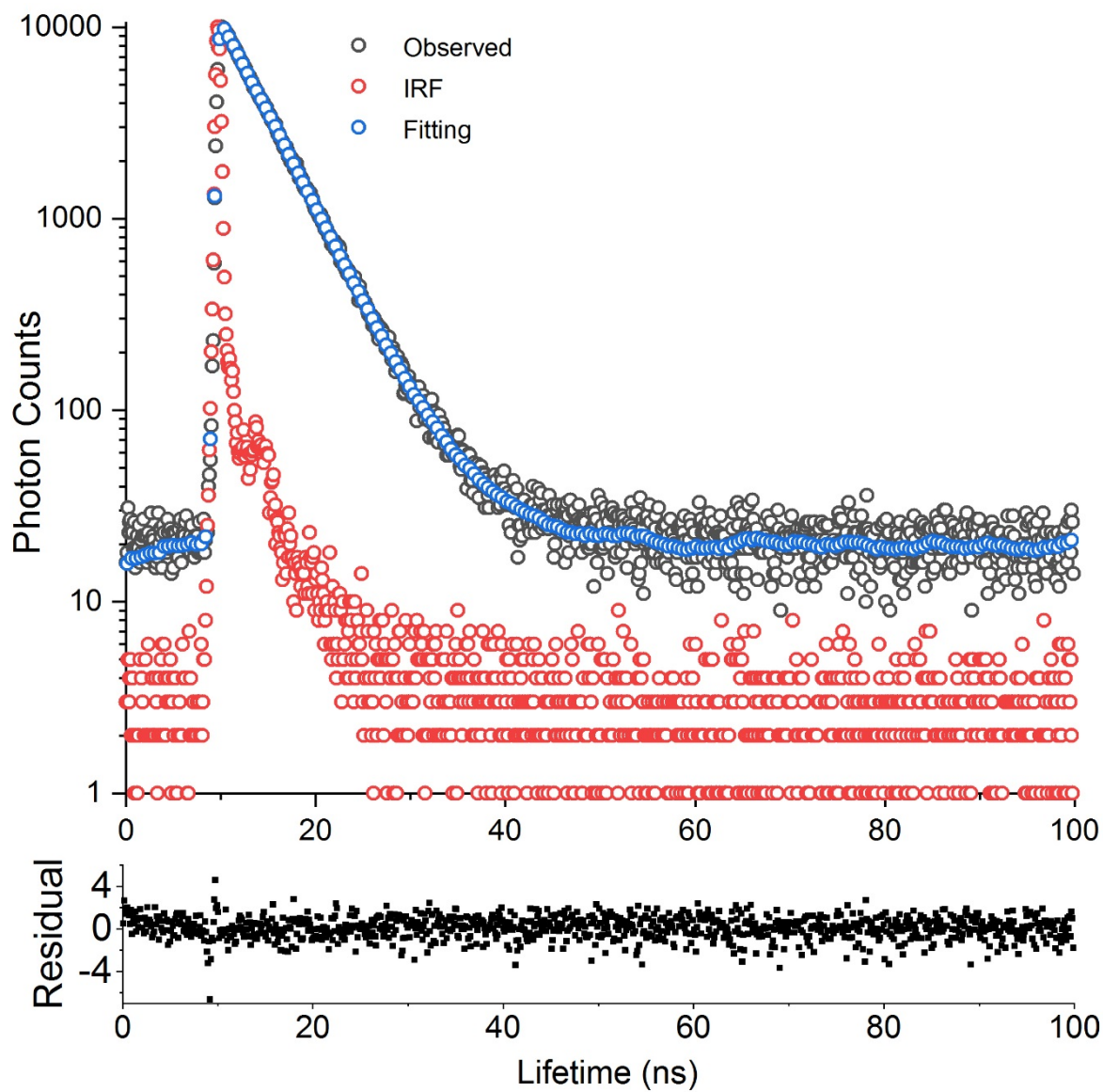

**Figure S17**. Lifetime decay profile of A+B solution (i.e. Rhodamine B and Fluorescein) measured for the emission maximum at 571 nm.

**Decay: A+B Solution@571nm emission**

    Lifetime Parameters:

| | |
|---|---|
| Time Range (ns) | : 0.00000 to 99.90234 step 0.09766 |
| Channel Range | : 0 to 1023 |
| Mode | : TCSPC |
| TAC (ns) | : 100 |
| Delay (ns) | : 0 |
| Time Calibration (ns) | : 0.09766 |
| Reps | : 1 |
| Acq Time (m:s) | : 01:36.5 |
| Ex Arm Parameters: | |
| WaveLength (nm) | : 362.50 |
| Bandwidth (nm) | : 0.01 |
| Lightpath | : TCSPC Diode |
| Em Arm Parameters: | |
| WaveLength (nm) | : 571.00 |
| Bandwidth (nm) | : 1.01 |

**IRF : A+B Solution@571nm emission**

| | |
|---|---|
| Lifetime Parameters: | |
| Time Range (ns) | : 0.00000 to 99.90234 step 0.09766 |
| Channel Range | : 0 to 1023 |
| Mode | : TCSPC |
| TAC (ns) | : 100 |
| Delay (ns) | : 0 |
| Time Calibration (ns) | : 0.09766 |
| Reps | : 1 |
| Acq Time (s) | : 11.3 |
| Ex Arm Parameters: | |
| WaveLength (nm) | : 362.5 |
| Bandwidth (nm) | : 0.01 |
| Lightpath | : TCSPC Diode |
| Em Arm Parameters: | |
| WaveLength (nm) | : 362.50 |
| Bandwidth (nm) | : 4.42 |